\documentclass[%
preprint,
 amsmath,amssymb,
 aps,
]{revtex4-2}

\usepackage{graphicx}
\usepackage{dcolumn}
\usepackage{bm}
\usepackage{subcaption}
\usepackage[shortlabels]{enumitem}
\usepackage{siunitx}
\graphicspath{{./figs/}}
\newcommand{\de}{^\circ}
\usepackage[normalem]{ulem}
\usepackage{comment}
\usepackage{tabularx}
\usepackage{array}

\usepackage[dvipsnames]{xcolor}
\newcommand{\red}{}

\begin{document}

\title{Analysis of dynamic stall development on a cross-flow turbine blade} 

\author{Mukul Dave}
  \email{mhdave@wisc.edu}
\author{Jennifer A. Franck}%
 \email{jafranck@wisc.edu}
\affiliation{%
 Department of Engineering Physics, University of Wisconsin-Madison,\\Madison, WI, USA - 53705.
}%

\date{\today}

\begin{abstract}
This research computationally investigates the complex dynamic stall phenomena of a cross-flow turbine blade utilizing modal analysis to identify pertinent events within the cycle. 
\red{The blade rotation perpendicular to the freestream generates a curved relative flow, a non-sinusoidal variation of relative flow speed and angle of attack, and the necessity of travelling through its own wake.
These complexities have challenged traditional predictors of dynamic stall such as pitch rate, pitching moment, or relative angle of attack.
To investigate these phenomena, aerodynamic loads and flow fields on the blade from large-eddy simulations are examined across two tip speed ratios.
Proper orthogonal decomposition of the velocity fields is employed to analyze the spatio-temporal evolution of the dominant flow features. The modes' time development coefficients reveal a stronger representation of the flow at the higher rotation rate, capturing the trend of relative flow velocity magnitude and lift generation on the blade, along with critical events such as vortex formation and detachment.
Additionally, mean power generation is enhanced by 40\% by applying a non-constant rotation rate (intracycle control or angular velocity control). The flow fields, supported by corresponding changes in the modal analysis, demonstrate that a delayed stall behavior is responsible for the additional power extraction.}
Finally, flow curvature, history effects, and \red{induced flow} are identified as significant factors that modify the dynamic stall onset and resulting force and moment curves as compared to \red{non-rotating} pitching or plunging foils. 
\end{abstract}

\maketitle


\section{\label{sec:intro}Introduction}
This research investigates the development of dynamic stall on a straight-bladed cross-flow turbine \red{over different kinematic regimes}. Flow over cross-flow turbines (CFT) is an active area of research due to their potential in harvesting \red{marine and} wind energy. However, a complete model of the fluid flow around CFT blades is lacking due to the cyclical variation of relative flow speed and angle of attack that induces flow separation even at optimal power producing rotation rates \cite{Snortland2019,Dave2021}. The periodic boundary layer separation and recovery, while typical of dynamic stall in other systems, is complicated by the rotation about the spanwise axis. The rotation produces Coriolis forces and \red{induced flow} that modify the nominal relative flow deduced from the kinematics \cite{Tsai2016}, changing the overall dynamics of the \red{flow} separation process. 

Past research has characterized the phases of dynamic stall on the basis of forces and moments acting on an airfoil in connection with temporal evolution of the velocity field and corresponding flow structures.
Carr \cite{Carr1988} reviews research on dynamic stall and the effect of different kinematic or flow parameters while providing a general chronology for the separation process and the aerodynamic load evolution.
The onset of stall is traditionally \red{delineated} by deviations in the force, suction pressure, or moment coefficients as reviewed by Sheng et al.~\cite{Sheng2006}. \red{Traditional dynamic stall onset may see the lift or normal force increasing beyond the static stall angle and a relatively constant moment coefficient followed by a rapid drop in the force and rapid increase in the pitch-down moment.}
In contrast to the explicit use of aerodynamic loads, Mulleners \& Raffel \cite{Mulleners2012} define the instance when the primary stall vortex detaches from the leading edge as an onset of its stalled phase. 
Proper orthogonal decomposition (POD) of the velocity field data from the experiments is used to identify the stall onset angle as the instance when the mode coefficient representing the dynamic stall vortex reaches a local maximum. 
Modal analysis of the surface pressure distribution from a pitching foil has also been employed to compare the flow trajectories across a parameter space \cite{Coleman2018} or to examine a bimodal distribution in cycle-to-cycle variations \cite{Ramasamy2020}.

A foil \red{with sinusoidal pitching kinematics is inadequate to understand stall on the rotating CFT blade due to the larger variations} in relative velocity magnitudes and angle of attack. A pitching foil that is simultaneously plunging, surging, or experiencing unsteady freestream flow is explored in literature for applications to helicopter blades \cite{Carta1979}, gusts \cite{Wong2013}, energy harvesting or propulsion \cite{Kim2017,Franck2017,Ribeiro2020,Dave2020}, and cross-flow turbines \cite{Dunne2015}. The plunging motion over a range of kinematic parameters delays the detachment or convection of the LEV and enhances lift \cite{Carta1979,Franck2017}. Gharali \& Johnson \cite{Gharali2013} show that a freestream varying out-of-phase to a pitching foil, as relevant to the cross-flow turbine, can result in lower loads than at static stall.
Furthermore, it is also demonstrated by Wong et al.~\cite{Wong2013} that a moving foil experiences a different force history than an unsteady inlet flow with identical relative velocity variation due to the fluid acceleration and the resulting distribution of vorticity at the trailing edge. Most relevant perhaps, is the work from Dunne \& McKeon \cite{Dunne2015} who extrapolate a low-order model using dynamic mode decomposition (DMD) modes from a pitching-surging foil to estimate the flow around a rotating CFT blade with similar relative velocity profiles.

\red{The on-blade flow physics of CFTs have also been investigated with a frequent emphasis on the formation and evolution of the stall vortex \cite{Fujisawa2001,SimaoFerreira2009}.
Due to their rotation perpendicular to the freestream flow, CFT blades experience distinct conditions from flow curvature and an undefined relative flow direction.}
A symmetric CFT blade profile with high chord-to-radius ratio behaves similar to a cambered foil in aligned flow with an incidence angle shift that effectively modifies the angle of attack computed geometrically at a single point \cite{Migliore1980,Bianchini2011,Balduzzi2014,Rainbird2015,Bianchini2016a}. 
Horst et al.~\cite{Horst2016} find that different airfoil transformation methods yield similar values of theoretically calculated virtual camber and incidence angle up to a chord-to-radius ratio of 0.4 but diverge beyond this limit. \red{
Furthermore, in flapping/surging foils, often only one side of the blade is investigated (due to symmetry), however the suction side of a CFT blade switches during the second half of its rotation.} Additionally, the blade encounters low relative velocity on the recovery portion of the stroke, which presumably influences the subsequent power stroke due to history effects. It is also found that Coriolis forces play a crucial role, particularly in determining the position of the stall vortex, which is not adequately modeled by a pitching-surging blade with identical relative flow \cite{Tsai2016}.

\red{Investigation of blade-level loads allows for comparison across different turbine geometries and operational conditions, as well as a comparison with dynamic stall research on non-rotating foils. Furthermore,}
modal analysis techniques such as POD or DMD that project the unsteady flow field onto a linear subspace are commonly used for analysis \cite{Mariappan2014,Mohan2016} or control strategies \cite{Seidel2005,Mohan2017,Taira2020} for dynamic stall.
With application to CFTs, while modal analysis has been used to examine the wake topology \cite{Strom2022,Scherl2022},
it has not been implemented at the blade level except by Le Fouest et al.~\cite{LeFouest2022} who developed a modal technique that automatically characterized the stall cycle across multiple rotation rates in lab experiments. This is presumably due to the difficulty of acquiring a full circumferential field of view around the blade throughout the rotation and the challenge posed by the moving reference frame in pre-processing of flow data.

\red{The aim of this paper is to utilize detailed large-eddy simulation (LES) to analyze dynamic stall development over two distinct regimes of angular velocity with potential application to flow control strategies such as active velocity control. Using POD modes of the velocity field, the resulting analysis directly compares the reduced-order representation of the flow with stall events and force trends throughout the cycle. Furthermore, the development and effect of these events are analyzed with respect to the changes in relative flow due to virtual camber and induced flow, which are unique to rotating foils. }

\red{Section \ref{sec:methods} describes the computational methodology, the modal analysis procedure, and details of the turbine geometry and kinematics. The rotational rate of the turbine, or tip speed ratio is varied from 1.9 to 1.1 to generate distinct stall regimes.  A third regime is investigated by imposing a sinusoidal angular velocity \cite{Strom2017}, improving turbine performance. 
Section \ref{sec:results} presents the dominant spatial modes and their time development coefficients for each simulation, which are analyzed in reference to the force or pitching moment on the foil and the evolution of surface pressure and vortices at the blade. Finally, section \ref{sec:discussion} discusses the implications of blade rotation and the effects it has on the dynamic stall process. }

\section{\label{sec:methods}Methods}

\subsection{Turbine kinematics}

\begin{figure}
    \centering
    \includegraphics[width=0.4\linewidth]{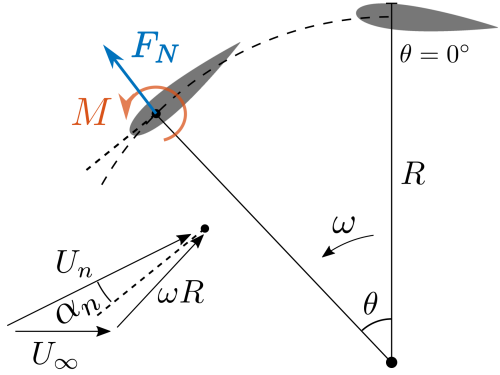}
    \caption{Kinematics and aerodynamic loads for the CFT blade.}
    \label{fig:schema}
\end{figure}

As shown in Fig.~\ref{fig:schema}, a CFT blade with chord length $c$ rotates in the freestream with an angular velocity of $\omega$ and a tip radius $R$. 
Lift is generated through the apparent angle of attack, $\alpha_n$, which is the angle the blade encounters with respect to the relative flow velocity, $\bm{U_n}$. The component of this lift force that is tangential to the blade motion drives the turbine.
Turbine performance is characterized as a function of tip speed ratio (TSR), defined as the ratio of turbine tip speed to the flow freestream velocity, $ \lambda(\theta) = {\omega(\theta)R}/{U_\infty} $.
In this paper, $\theta=0^\circ$ represents the foil position when the foil velocity is directly opposite the freestream velocity.
The sign of $\alpha_n$ is positive when the relative flow velocity is directed outwards from the axis of rotation and negative when it is directed inwards as in Fig.~\ref{fig:schema}. 

\subsection{Turbine performance and blade loading}

The torque on the blade about its center of rotation, $q$, and the resulting power being generated are normalized as the torque coefficient and power coefficient given by
\begin{equation*}
    C_Q(\theta) = \frac{q(\theta)}{\frac{1}{2}\rho\ U_\infty^2 AR}\ \ \text{and} \ \ 
    C_P(\theta) = \frac{q(\theta) \omega(\theta)}{\frac{1}{2}\rho\ U_\infty^3A}\ ,
\end{equation*}
where $A$ is projected area of the turbine normal to the freestream. The blade aerodynamic loads that are analyzed for characterizing stall are the normal force, \red{$F_N$}, and the pitching moment about the quarter-chord location, $M$, as shown in Fig.~\ref{fig:schema}. 
The loads are normalized using the blade velocity, $\omega R$, as the pitching moment coefficient and the normal force coefficient given by
\begin{equation*}
    \red{C_M}(\theta) = \frac{M(\theta)}{\frac{1}{2}\rho\ (\omega(\theta) R)^2\ Sc}\ \ \text{and} \ \ 
    \red{C_N}(\theta) = \frac{\red{F_N}(\theta)}{\frac{1}{2}\rho\ (\omega(\theta) R)^2\ S}\ ,
\end{equation*}
where $S$ is planform area of the blade. 
The use of blade velocity for normalization allows a quantitative comparison of the loads between \red{simulations of different tip speed ratios}.
The moment, $M$, is positive in the counter-clockwise direction and \red{$F_N$} is positive when acting outward from the turbine as demonstrated in Fig.~\ref{fig:schema}. To draw connections with vortex structures, the surface pressure is computed along the blade and is normalized by the freestream pressure $12R$ upstream from the turbine center.

\subsection{Turbine geometric and kinematic parameters}

The turbine has two straight-bladed NACA0018 foils with $c/R=0.47$ and a preset pitch angle of $\alpha_p = 6^\circ$ (leading edge angled outward). 
The Reynolds number is $Re = {c U_\infty}/{\nu} = 4.5\times10^4$ and the blockage ratio, defined as the ratio of the turbine cross-section to that of the flow, is 10.6\%.
\red{The turbine geometry and flow conditions including Reynolds number and blockage match flume experiments \cite{Snortland2019} with the exceptions that the blades have infinite span, no support structure is modeled, and the turbulent intensity of the oncoming flow is not matched.
Although a field turbine is expected to operate at a higher $Re$ with added geometrical and flow complexities, the current simulations at a moderate $Re$ allow for a wall-resolved and validated numerical analysis of the complex dynamic stall physics on the blade.}
The LES has shown that while the confinement level of 10.6\% increases power generation, it does not alter the stall cycle qualitatively compared to the unconfined simulation \cite{Dave2021}.

\begin{center}
\begin{table}
\begin{tabular}{ m{8em} m{13em} m{15em} m{3em}}
 \hline
 Simulation  & Tip speed ratio & Rationale & $\overline{C_P}$ \\
 \hline 
 optimal TSR & $\lambda=1.9$ & Peak power for constant TSR & 0.284 \\ 
 sub-optimal TSR & $\lambda=1.1$ & High angles of attack/deep stall & 0.081 \\ 
 sinusoidal TSR & $\overline{\lambda}=1.9$, $A_{\lambda}=1.3\  \overline{\lambda}$, $\phi_{\lambda}=4$ & Optimized for power & 0.396 \\ 
 \hline
\end{tabular}
\caption{\red{Simulations performed with tip speed ratio, power coefficient from the two blades, and rationale for investigation.} }
\label{table:1}
\end{table}
\end{center}

\red{Table \ref{table:1} lists the three turbine simulations performed along with the kinematics and mean power coefficients.}
A constant angular velocity rotation with $\lambda=1.9$ is explored, which represents the optimal \red{TSR} for this turbine configuration as found in prior experiments \cite{Snortland2019}. The optimal constant TSR experiences flow separation and hence investigating its stall cycle is of particular interest for design and control purpose. 
Next, \red{a sub-optimal TSR of} $\lambda=1.1$ with angles of attack up to $60^\circ$ is simulated. It undergoes drastic flow separation and hence acts as a limiting case to test the hypotheses correlating flow physics with blade loads and POD modes. 

Rather than rotating the turbine with a constant angular velocity, the rotation rate can be varied cyclically with an aim to control the magnitude and direction of the relative flow and hence the stall cycle, which is referred to as \textit{intracycle} variation.
An optimized sinusoidal variation of angular velocity \red{or sinusoidal TSR} is found to enhance power generation by up to 53\% in experiments by Strom et al.~\cite{Strom2017}.
To understand the underlying modifications in stall dynamics, \red{the sinusoidal TSR is simulated as
\begin{equation}
\lambda(\theta) = \overline{\lambda} + A_\lambda sin(2\theta + \phi_\lambda) ,
\label{eqn:omega_intra}
\end{equation}
where the mean, $\overline{\lambda}=1.9$, $A_\lambda=1.3\ \overline{\lambda}$, and the phase shift, $\phi_\lambda= \SI{4.0}{\radian}$, are set close to the optimal values \cite{Strom2017}.}

\subsection{Simulation setup}
Large-eddy simulations (LES) of the cross-flow turbine are performed that solve the spatially filtered Navier-Stokes equations while using a model to account for the smallest (sub-grid) length scales. 
The dynamic $k$-equation model is employed which solves a transport equation for the sub-grid scale kinetic energy and computes the model coefficients dynamically from local flow properties.
The boundary layer on the blade is fully resolved and a periodic condition is implemented along the span to model an infinitely long blade with no tip effects.
A mesh and model sensitivity analysis along with experimental validation has been previously performed for this setup \cite{Dave2021}. 
The chosen mesh has 48 extruded layers over a blade span of length $0.2c$, with a total of 4.54 million cells.

Ten rotations of the turbine are simulated for optimal TSR and the sinusoidal TSR, and nine rotations are simulated for sub-optimal TSR. \red{LES simulations previously published have investigated the transient flow and the cycle-to-cycle variations of the turbine blade and found that the power cycle converges after four rotations with a RMS difference value of less than 0.05 (see \cite{Dave2021} for more details), indicating that the dominant flow physics on the blade are not significantly changing from one rotation to the next. Thus, for the modal analysis below, the first four cycles are eliminated leaving five or six cycles for analysis. Considering the flow in a frame of reference of the blade for a two-bladed turbine, this provides 10-12 unique blade rotations.} 
Flow data at 70 azimuthal positions per rotation for optimal TSR, 90 for sinusoidal TSR, and 121 for sub-optimal TSR provide a total of 840, 1080, and 1210 snapshots respectively from the two blades with uniform time intervals. 

\subsection{Proper orthogonal decomposition}
\begin{figure}
    \centering
    \includegraphics[width=\linewidth]{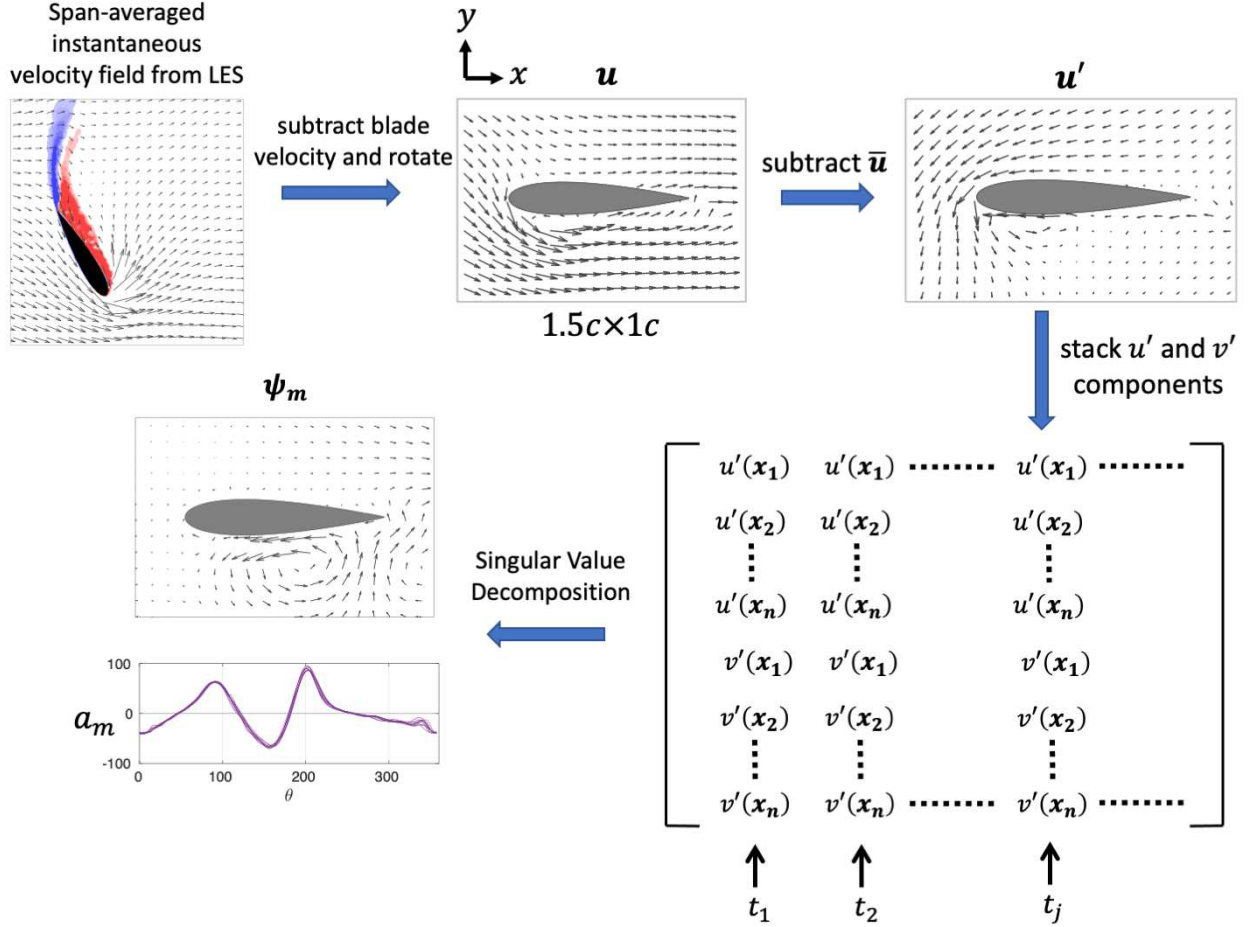}
    \caption{Schematic of data processing of velocity fields for POD analysis.}
    \label{fig:pod}
\end{figure}

A low-dimensional basis is constructed for the full circumferential flow field around the rotating CFT blade through proper orthogonal decomposition (POD) implemented on flow data from LES.
Span-averaged velocity, normalized by the freestream velocity, is extracted on a grid of resolution $0.01c$ from a region of size $1.5c \times 1c$ centered at the blade for optimal TSR and sinusoidal TSR. For sub-optimal TSR, the field of view is expanded to $2c \times 1.5c$ to include more area on the inside of the blade containing the large stall vortex.
To provide a uniform reference frame across time, a velocity field relative to the blade, $\bm{u}$, is calculated by subtracting the instantaneous blade velocity and the blade position is fixed horizontally\red{, shown schematically in Fig.~\ref{fig:pod}.} 
Next, the time-averaged mean velocity field is subtracted from all snapshots to obtain the fluctuating component, $\bm{u'} = \bm{u} - \bar{\bm{u}}$. Velocities along $x$-direction (along the blade) and $y$-direction (lateral to the blade) on the two-dimensional grid from each time snapshot are then stacked as a one-dimensional column vector, with time advancement along rows of the matrix, known as the method of snapshots \cite{Sirovich1987}. \red{Finally, a singular value decomposition of this matrix provides discrete vectors that are interpreted as spatial modes, $\bm{\psi_m}(x,y)$, and their time development coefficients, $a_m(t_j)$.} The flow can be reconstructed as $\bm{u'}(x,y,t_j) = \sum_{m=1}^{N} a_m(t_j) \bm{\psi_m}(x,y)$, where $m$ is the mode index in decreasing order of their singular values or significance, and $j$ is the time snapshot index. The mode coefficients are presented in terms of the azimuthal position, $\theta_j$.

\section{\label{sec:results}Results}

\subsection{Optimal tip speed ratio, $\lambda=1.9$}

For the geometric configuration and Reynolds number regime in this research, $\lambda=1.9$ represents the optimal constant tip speed ratio (TSR) for energy generation.
Whereas the torque generation and turbine-level flow field were previously analyzed qualitatively by the authors \cite{Dave2021}, this research focuses on a blade level analysis of the stall cycle by associating the flow physics with surface pressure, normal force, pitching moment, and POD modes.

\begin{figure}
    \captionsetup[subfigure]{font={footnotesize,stretch=0.9}}
    \centering
    \begin{minipage}{0.57\textwidth}%
        \begin{subfigure}{\linewidth}%
            \includegraphics[width=\linewidth]{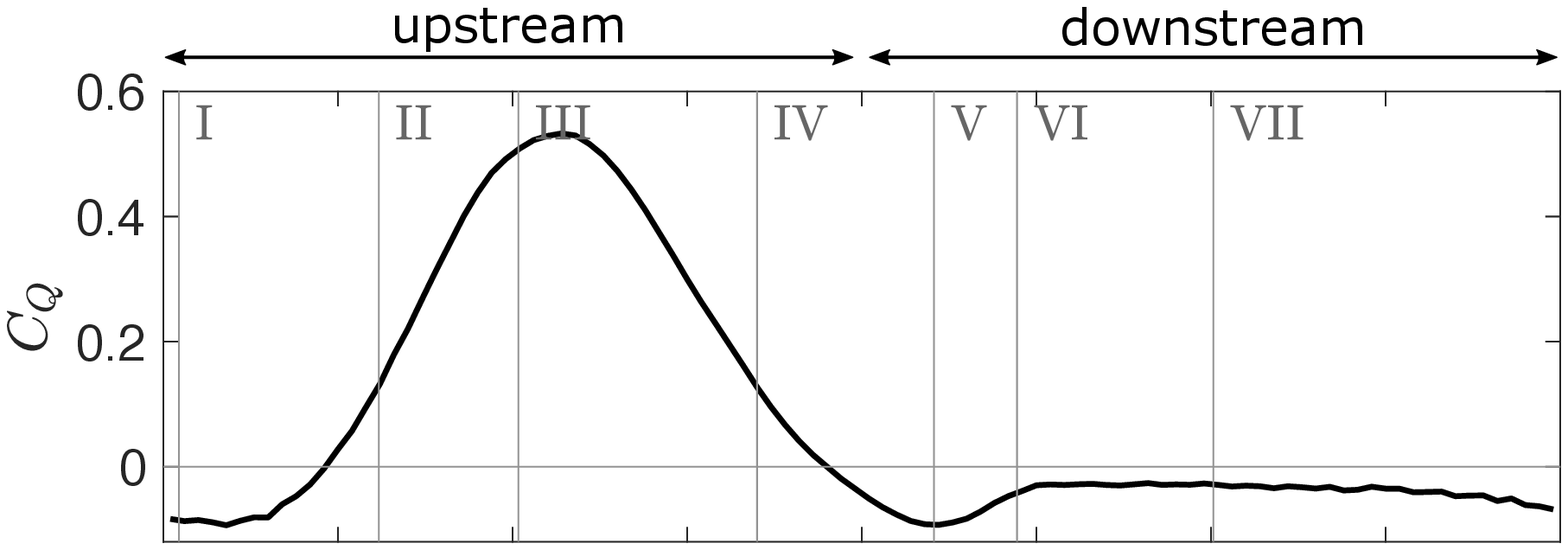}
            \caption{Torque coefficient} 
            \label{fig:const_all_1_9_a}
        \end{subfigure}
        \begin{subfigure}{\linewidth}%
            \includegraphics[width=\linewidth]{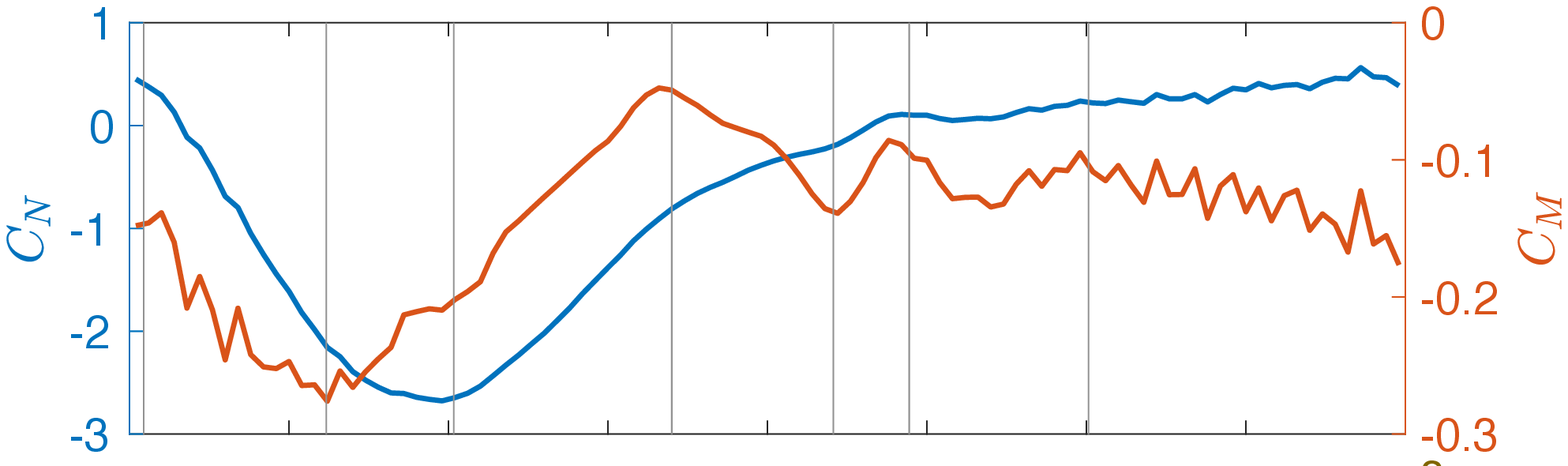}
            \caption{Normal force (blue; left axis) and \protect\linebreak pitching moment (red; right axis)} 
            \label{fig:const_all_1_9_b}
        \end{subfigure}
        \begin{subfigure}{\textwidth}%
            \includegraphics[width=\linewidth]{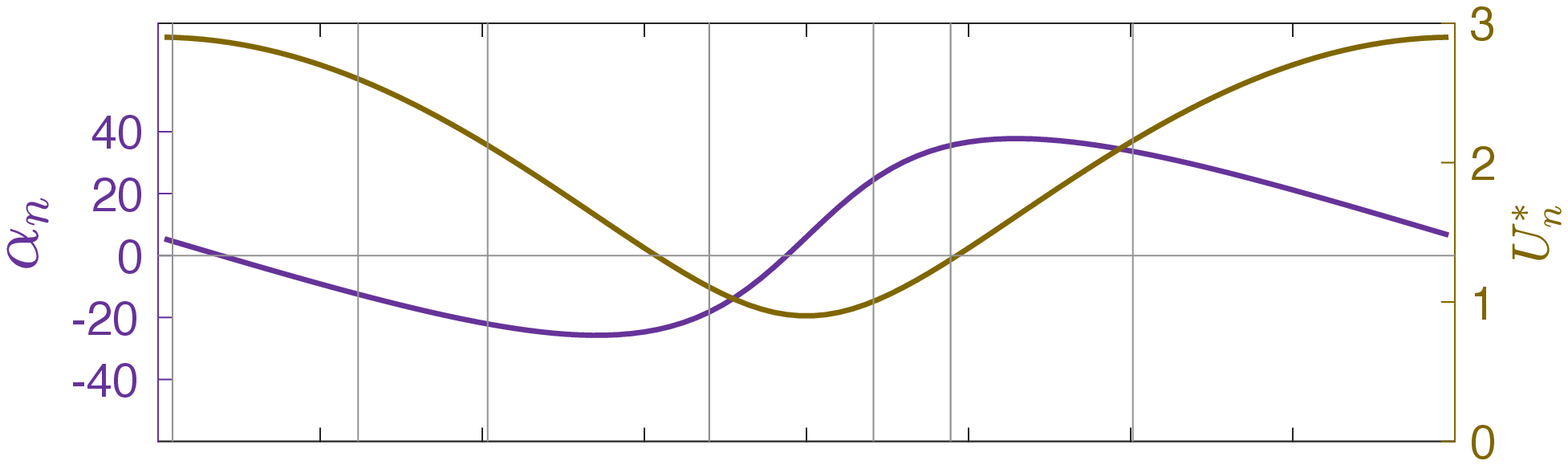}
            \caption{Nominal angle of attack (purple; left axis) \protect\linebreak and relative flow velocity (brown; right axis)} 
            \label{fig:const_all_1_9_c}
        \end{subfigure}
        \begin{subfigure}{\textwidth}%
            \includegraphics[width=\linewidth]{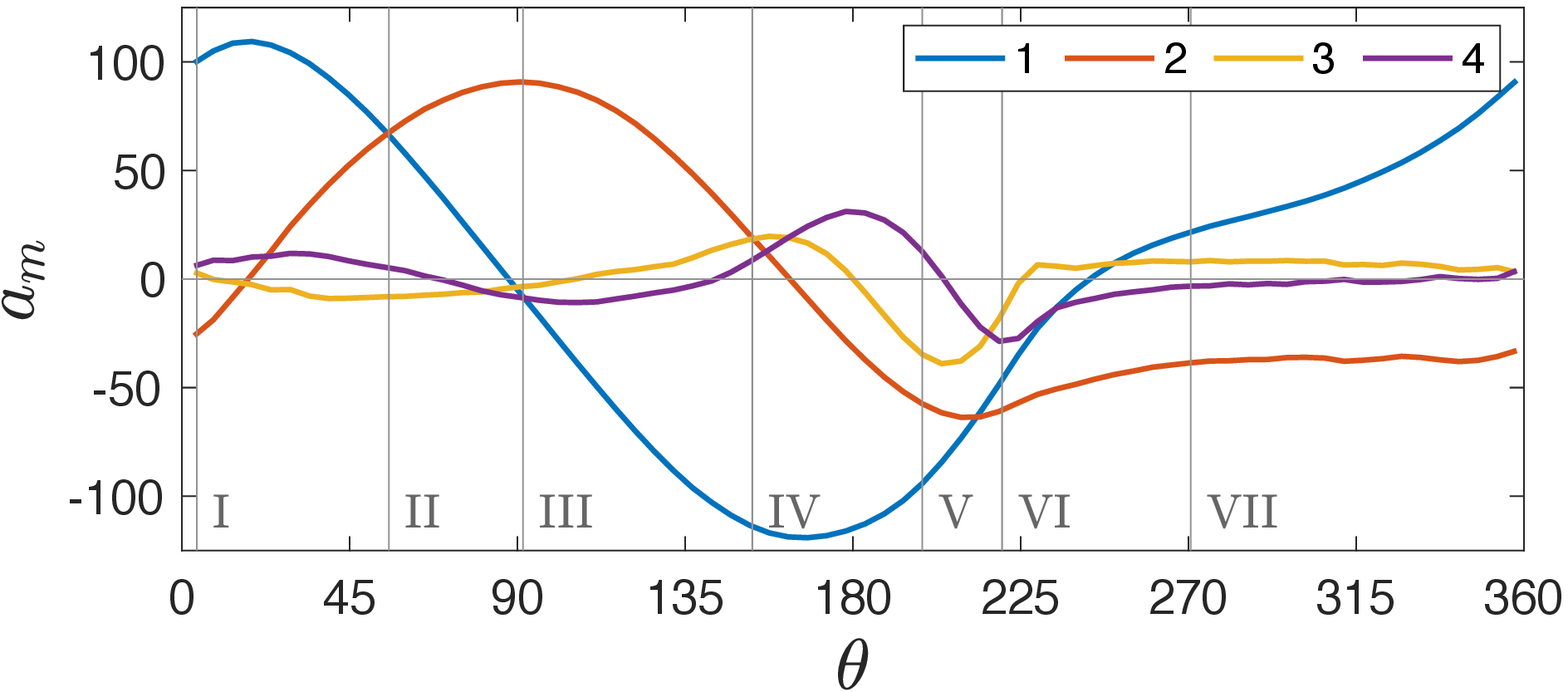}
            \caption{Time development coefficients: modes 1-4} 
            \label{fig:const_all_1_9_d}
        \end{subfigure}
    \end{minipage}
    \hfill
    \begin{minipage}{0.42\textwidth}%
    \begin{subfigure}{\linewidth}%
        \includegraphics[width=\textwidth]{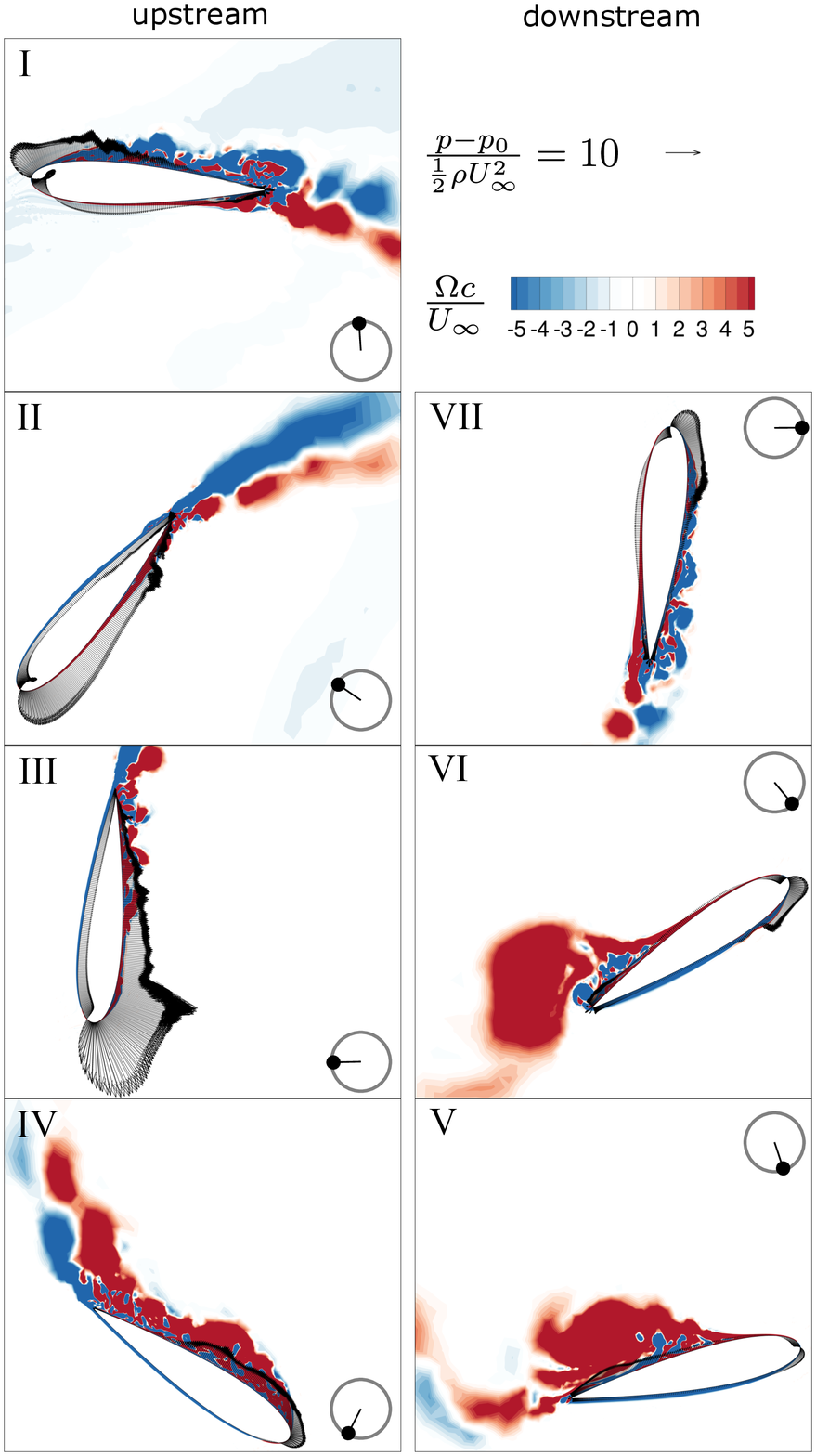}
        \caption{Instantaneous vorticity contours superimposed with surface pressure vectors for specific blade positions (I-VII); reference scale for pressure vectors in upper right.}
        \label{fig:const_all_1_9_e}
    \end{subfigure}
    \end{minipage}%
    \caption{Analysis through a single blade rotation for optimal TSR.}
    \label{fig:const_all_1_9}
\end{figure}

The kinematics and dynamics of the blade, including the nominal relative flow metrics $\alpha_n$ and \red{$U_n^*= U_n/U_\infty$}, the instantaneous aerodynamic loads \red{$C_N$ and $C_M$}, and the torque $C_Q$ are displayed in Fig.~\ref{fig:const_all_1_9}.
The nominal relative flow is calculated geometrically from the kinematics and is expected to differ from true relative flow due to \red{induced flow such as} reduced velocity on the downstream or recovery portion of the stroke.
The pitching moment is negative in the clockwise direction, implying a ``pitch-down" moment relative to the angle of attack during the upstream rotation of the turbine, and a ``pitch-up" moment relative to the nominal angle of attack during the downstream rotation.

To visualize critical points within the cycle, the instantaneous vorticity fields are shown in Fig.~\ref{fig:const_all_1_9}. Instantaneous surface pressure is also displayed as a series of vectors acting along the blade surface normal, where a vector acting outward from the blade implies suction on the surface.

At the start of the upstream motion in phase I of Fig.~\ref{fig:const_all_1_9}, the nominal angle of attack is slightly positive.
The blade has a small outward normal force due to the partial flow separation on its outer surface during its downstream rotation which also causes a \red{``pitch-up"} moment. 
    
Between phases I-II, a suction pressure develops on the inside of the blade due to the increasing angle of attack, while the flow reattaches on the outside. At II, the clockwise (CW) pitching moment has peaked and the flow separates on a small portion near the trailing edge.   
A positive pressure develops on the outside of the blade in phase II and III due to reattachment of the boundary layer. 
This tends to enhance the inward normal force and the clockwise tendency of the pitching moment.
However, the flow separation on the inside 
at phase III results in loss of suction pressure and a counter-clockwise (CCW) pitching moment from the suction at the leading edge, reducing the overall CW moment. The normal force reaches a peak at phase III due to flow separation and a reducing value of $U_n$. 

During phase III-IV, the positive pressure on the outside of the blade subsides as the flow is fully reattached. On the inside of the blade, the separation point moves towards the leading edge and subsequently an LEV forms, concentrating the suction pressure. 
The onset of stall occurs after phase IV when the LEV detaches and convects towards the trailing edge, thus increasing the CW pitching moment. The nominal angle of attack rapidly switches side as the LEV convects along the blade in Fig.~\ref{fig:const_all_1_9}, phase V. A local peak occurs in the CW moment and torque due to the corresponding suction pressure.

On the downstream \red{portion of the stroke}, the oncoming flow (see velocity vector diagram in Fig.~\ref{fig:schema} for reference) is smaller and hence the angles of attack are lower than the nominal values plotted in Fig.~\ref{fig:const_all_1_9_c}.
This prevents fully separated flow but is adequate to cause separation and hence loss of suction pressure on the trailing portion of the outer surface. 
A suction pressure develops on the outside leading edge by phase VI causing the pitching moment to remain in the CW direction but the normal force is now acting outward from the turbine center.
At phase VII, partial flow separation is observed on the outside surface with attached flow and strong suction pressure at the leading edge, resulting in a persistent CW pitching moment and a low outward normal force. 

\begin{figure}
    \centering
    \begin{subfigure}{\columnwidth}
        \centering
        \includegraphics[width=\linewidth]{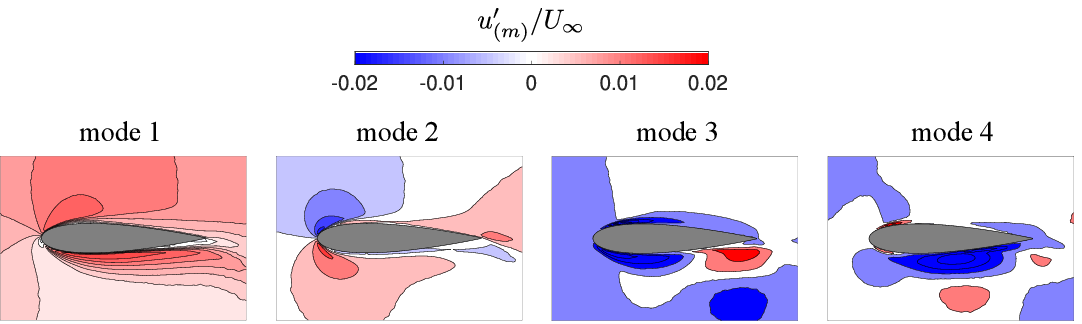}
        \caption{$x$-velocity fields}
    \end{subfigure} 
    
    \begin{subfigure}{\columnwidth}
        \centering
        \includegraphics[width=\linewidth]{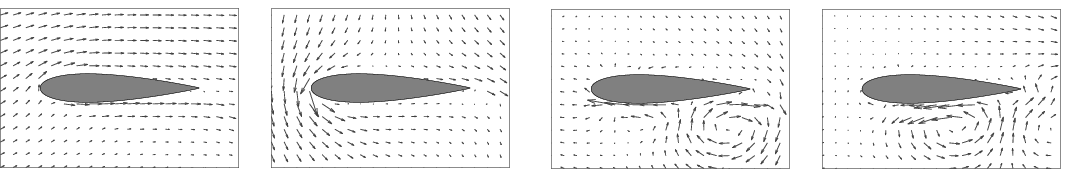}
        \caption{\red{2D velocity vectors}}
    \end{subfigure}

    \caption{Visualization of first four modes for optimal TSR ($\lambda=1.9$).}
    \label{fig:const_modes_u}
\end{figure}

\red{Fig.~\ref{fig:const_modes_u} presents the first four modes from POD in the blade referenced coordinate system, capturing 95.6\% of the energy. The $x$-direction aligns with the chord, and the bottom of the foil corresponds to inside of the rotating blade, facing the turbine center.
The velocity vectors in Fig.~\ref{fig:const_modes_u}b illustrate the entire mode whereas the $x$-component in Fig.~\ref{fig:const_modes_u}a demonstrates changes around the foil with a higher spatial resolution.}
The respective time development coefficients of the four modes are displayed in Fig.~\ref{fig:const_all_1_9}d.

Mode 1 displays attached flow along the foil with a minimal angle of attack and hence its temporal coefficient closely imitates the relative velocity magnitude $U_n^*$, shown in Fig.~\ref{fig:const_all_1_9}c, on the upstream stroke from $\theta=0^\circ$ to $180^\circ$. The asymmetry of the time development coefficients between the upstream ($\theta=0^\circ$ to $180^\circ$) and downstream ($\theta=180^\circ$ to $360^\circ$) side are due to the discrepancy between nominal and actual relative flow experienced on the downstream stroke.

Mode 2 depicts the flow sharply bending around the leading edge, and the trailing edge to a lesser extent, indicating a velocity differential over opposite sides that is typical of lift generation.  
Its coefficient aligns with the torque generation by the blade in Fig.~\ref{fig:const_all_1_9}a since both peak around $\theta=95^\circ$ and have a negative peak around $\theta=215^\circ$.

Modes 3 and 4 prominently include large regions of reverse flow at the blade surface along with vortical flow as observed by neighboring regions of opposite signed velocities. The positive and negative peaks correspond to the leading edge vortex formation and shedding. For instance, the peak of mode 3 at phase IV represents the onset of stall, ultimately leading to position VI that marks the negative peak of mode 4 and the instance when the vortex has passed the trailing edge.


\subsection{Sub-optimal tip speed ratio, $\lambda=1.1$}

\begin{figure}
    \captionsetup[subfigure]{font={footnotesize,stretch=0.9}}
    \centering
    \begin{minipage}{0.57\textwidth}%
        \begin{subfigure}{\linewidth}%
            \includegraphics[width=\linewidth]{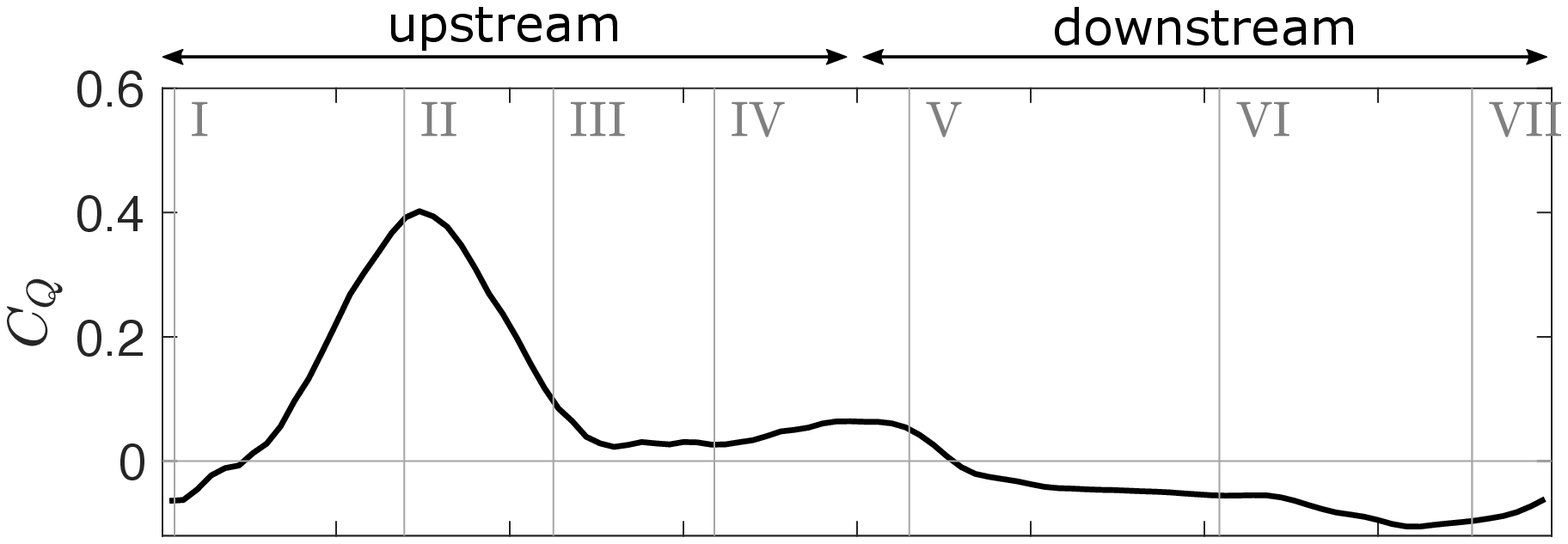}
            \caption{Torque coefficient} 
            \label{fig:const_all_1_1_a}
        \end{subfigure}
        \begin{subfigure}{\linewidth}%
            \includegraphics[width=\linewidth]{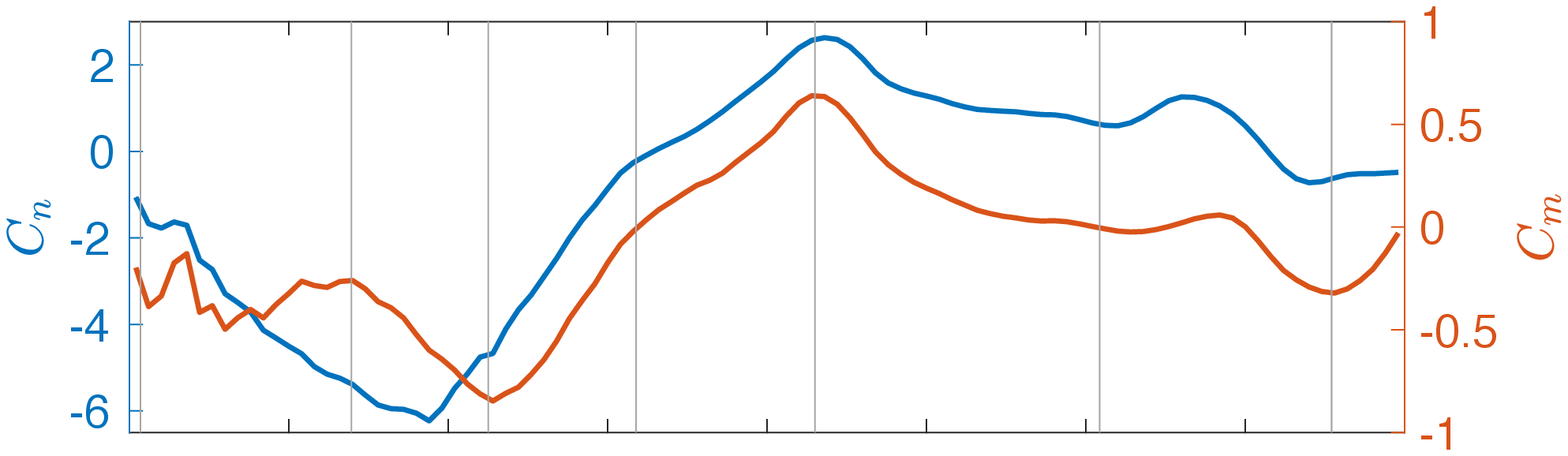}
            \caption{Normal force (blue; left axis) and \protect\linebreak pitching moment (red; right axis)} 
            \label{fig:const_all_1_1_b}
        \end{subfigure}
        \begin{subfigure}{\textwidth}%
            \includegraphics[width=\linewidth]{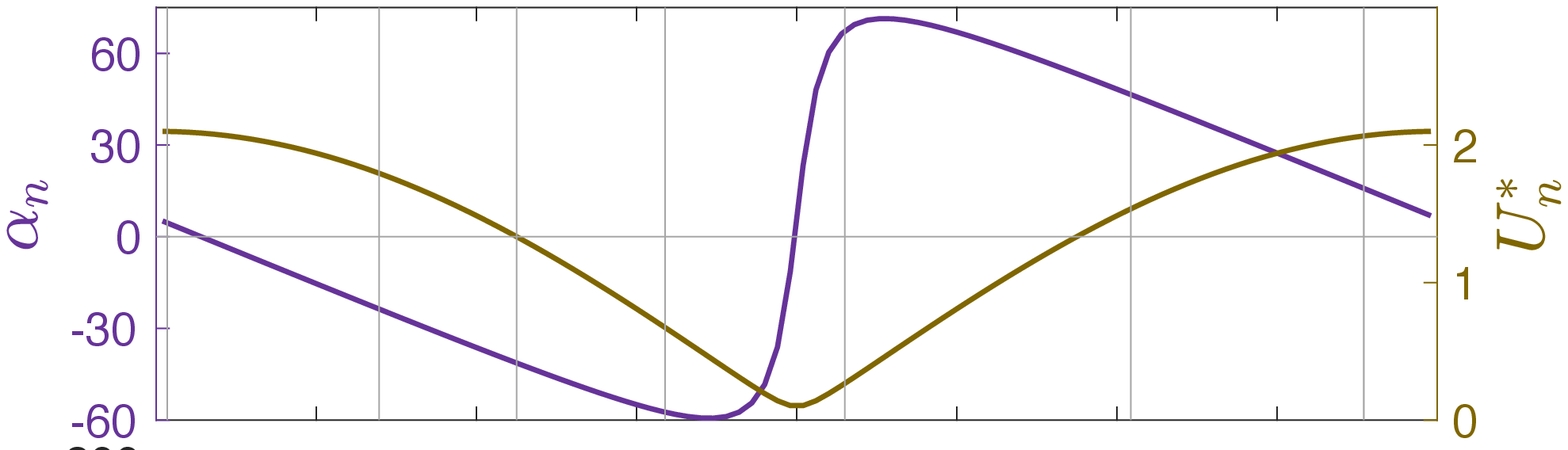}
            \caption{Nominal angle of attack (purple; left axis) \protect\linebreak and relative flow velocity (brown; right axis)} 
            \label{fig:const_all_1_1_c}
        \end{subfigure}
        \begin{subfigure}{\textwidth}%
            \includegraphics[width=\linewidth]{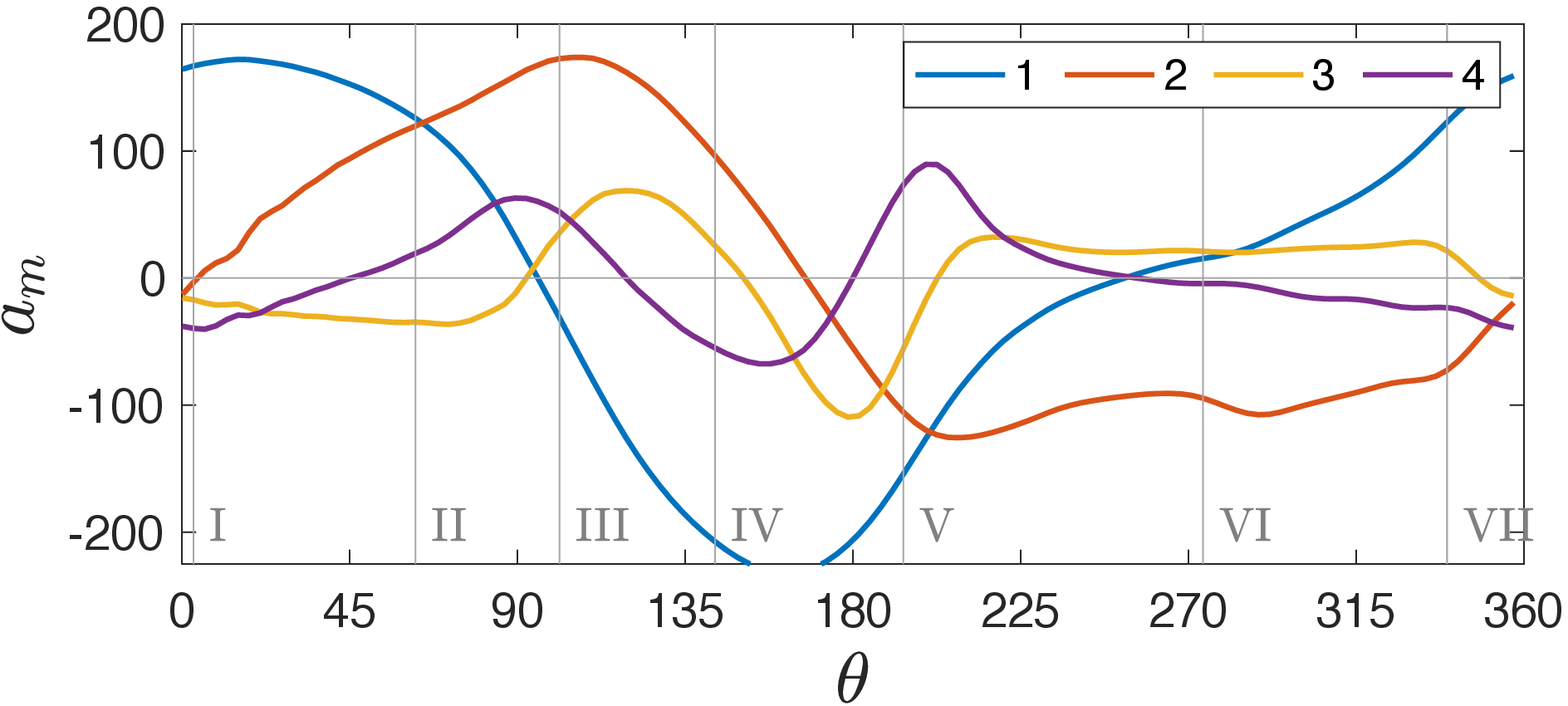}
            \caption{Time development coefficients: modes 1-4} 
            \label{fig:const_all_1_1_d}
        \end{subfigure}
    \end{minipage}
    \hfill
    \begin{minipage}{0.42\textwidth}%
    \begin{subfigure}{\linewidth}%
        \includegraphics[width=\textwidth]{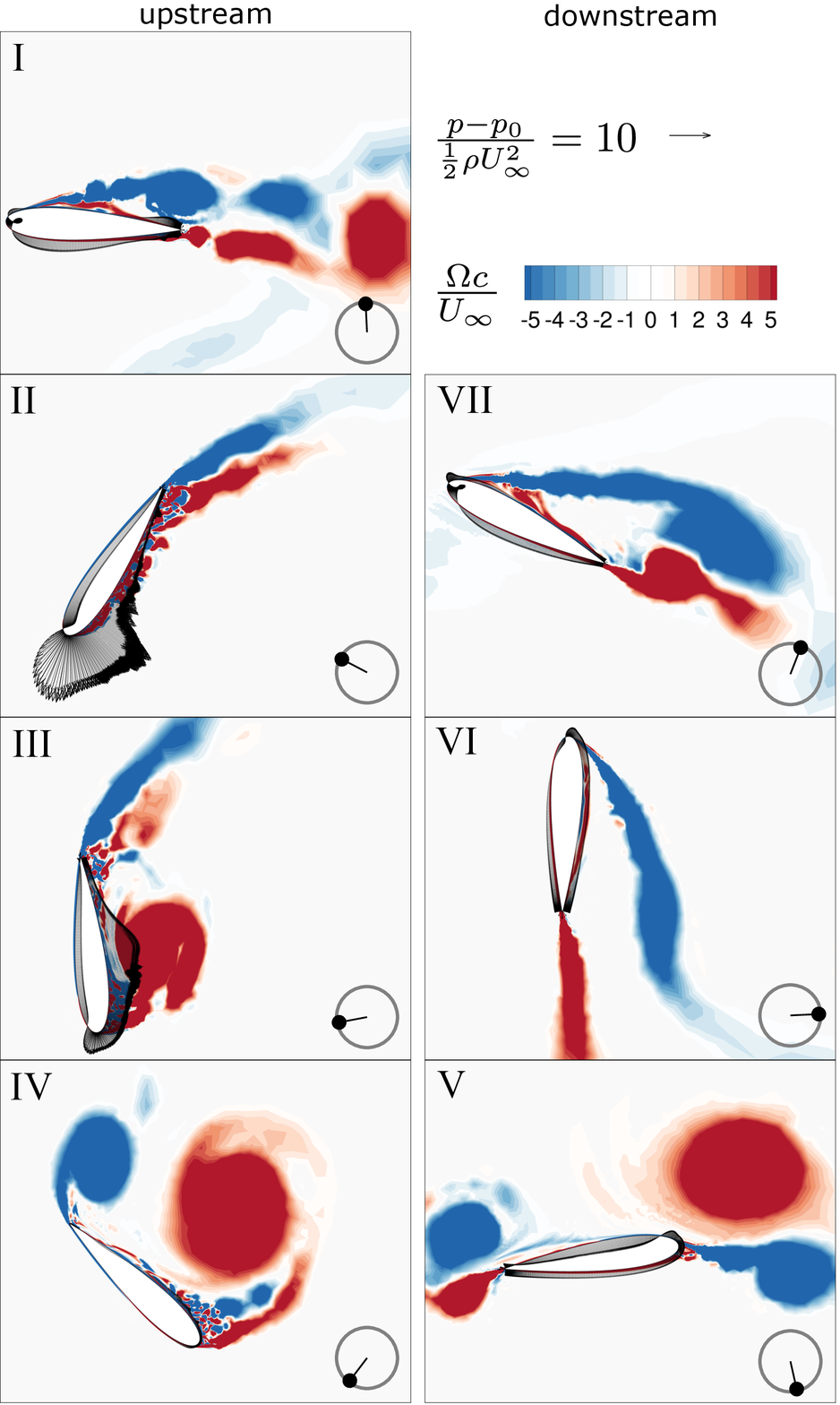}
        \caption{Instantaneous vorticity contours superimposed with surface pressure vectors for specific blade positions (I-VII); reference scale for pressure vectors in upper right.}
        \label{fig:const_all_1_1_e}
    \end{subfigure}
    \end{minipage}%
    \caption{Analysis through a single blade rotation for sub-optimal TSR.}
    \label{fig:const_all_1_1}
\end{figure}

A tip speed ratio of $\lambda=1.1$ is simulated to explore the physics and the effectiveness of POD in capturing the dominant features within a deep stall regime. 
Within this regime, the steeper variations in nominal angles of attack (up to $60\de$) and relative velocity produce stronger force and moment fluctuations, and more complex vortex dynamics. The torque coefficient, aerodynamic loads on the blade, relative blade kinematics, and flow fields are shown in Fig.~\ref{fig:const_all_1_1} for direct comparison with optimal TSR in Fig.~\ref{fig:const_all_1_9}.

At the top of the stroke, in contrast with the partial flow separation for optimal TSR, significant separation and shedding occurs in phase I. Although the positive nominal angle of attack is identical to that for the higher tip speed ratio, the effective angle of attack is likely lower, as observed by the location of stagnation pressure along the leading edge. A suction pressure is beginning to form along the inside of the blade that develops into a small fluctuating CW moment.
As the relative angle of attack dramatically increases in magnitude, the outside boundary layer has reattached in phase II with partial separation along the inside of the blade. Compared to the optimal TSR flow, development of the CW pitching moment is delayed, increasing between II and III as the flow separates at the leading edge and immediately rolls up into a larger stall vortex.  

The LEV detaches and induces an opposite signed trailing edge vortex at IV as the CW moment subsides.
As observed at phase V, the vortex convects upstream of the leading edge due to lower relative flow velocities and smaller Coriolis forces with respect to the optimal TSR flow. 
Since the nominal relative flow velocity is close to zero, the relative flow observed by the blade is dominated by the leading and trailing edge vortices which are inducing a flow normal to the blade creating an outward normal force and a CCW pitching moment. 
Additionally, two weaker vortices are formed at the leading and trailing edges on the outside of the blade due to the flow induced by the larger vortices. 
As the relative velocity increases from V-VI and the vortices are convected downstream of the foil, a fully separated flow develops.
The shear layer for the separating flow at VI indicates the reduction in angle of attack from VI-VII along with a periodic vortex shedding. The flow continues to be separated until the angle of attack switches side again after phase I.

\begin{figure}
    \centering
    \includegraphics[width=\linewidth]{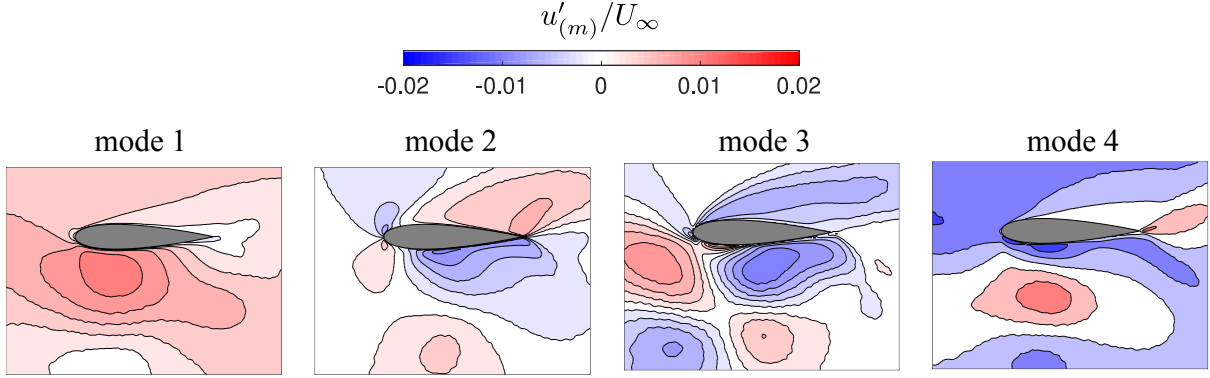}
    \caption{Visualization of first four modes with $x$-velocity contours for sub-optimal TSR ($\lambda=1.1$).}
    \label{fig:const11_modes_u}
\end{figure}

\red{The first four modes of the POD analysis are shown in Fig.~\ref{fig:const11_modes_u} with corresponding time development coefficients in Fig.~\ref{fig:const_all_1_1_c}. The modes' shape differ from Fig.~\ref{fig:const_modes_u} due to the larger degree of flow separation across the entire cycle. Mode 1 represents low angle of attack flow similar to optimal TSR, however with an additional vortex structure on the inside. Mode 2 illustrates flow almost orthogonal to the chord, bending more prominently around the trailing edge as observed for separated flows. As shown in Fig.~\ref{fig:const_all_1_1_d}, mode 2 surpasses mode 1 at phase II, coincident with maximum torque, then peaks at phase III at the maximum CW moment. The mode 2 trend is similar to that in optimal TSR, except that it is combined with modes 3 and 4 earlier in the cycle.}

\red{Modes 3 and 4 account for the LEV and its convection from the blade, which demonstrate more complexity of the vortex dynamics in contrast with the light stall at optimal TSR. At phase III, the vortex is centered closer to mid-chord (rather than the leading edge for optimal TSR) and corresponds to peaks in mode 4 and mode 3 even though mode 2 continues to be activated. As the foil enters the downstream stroke, all four mode coefficients play a role in describing the flow. For the sub-optimal TSR, the four modes only capture 90.6\% of the energy, lower than the optimal TSR case. The POD reconstruction with eight modes captures 96.5\% of the energy, with modes 5-8 capturing the higher order vortex dynamics.}

\subsection{\label{sec:sine}Sinusoidal tip speed ratio}

\begin{figure}
    \captionsetup[subfigure]{font={footnotesize,stretch=0.9}}
    \centering
    \begin{minipage}{0.49\textwidth}%
        \begin{subfigure}{\linewidth}%
            \includegraphics[width=\linewidth]{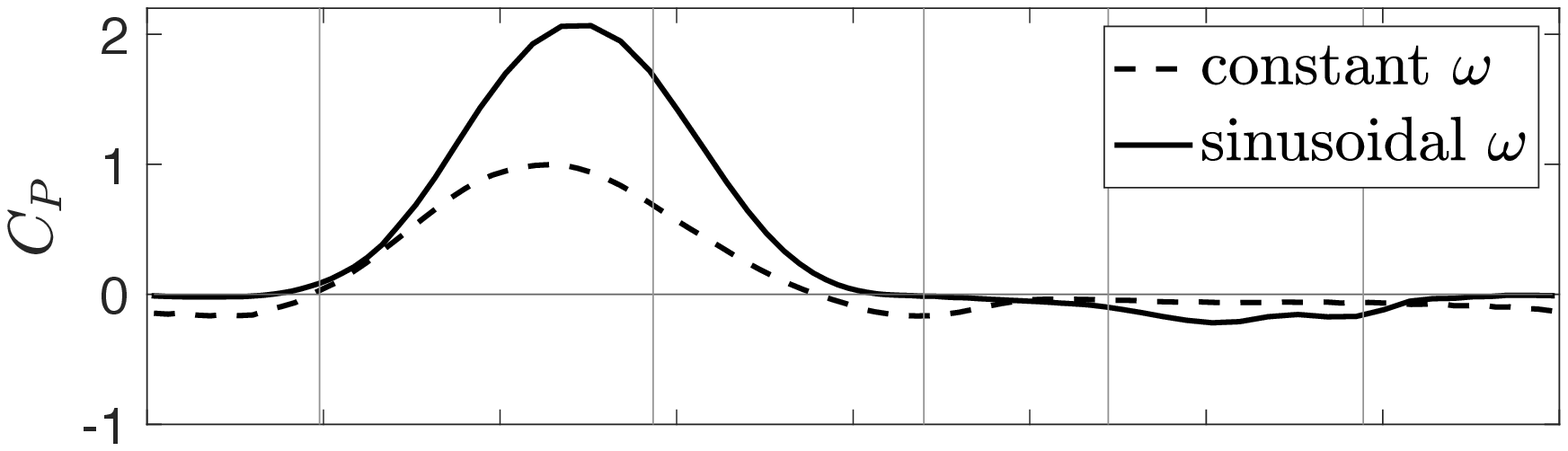}
            \caption{Power coefficient} 
            \label{fig:intra_a}
        \end{subfigure}
        \begin{subfigure}{\linewidth}%
            \includegraphics[width=\linewidth]{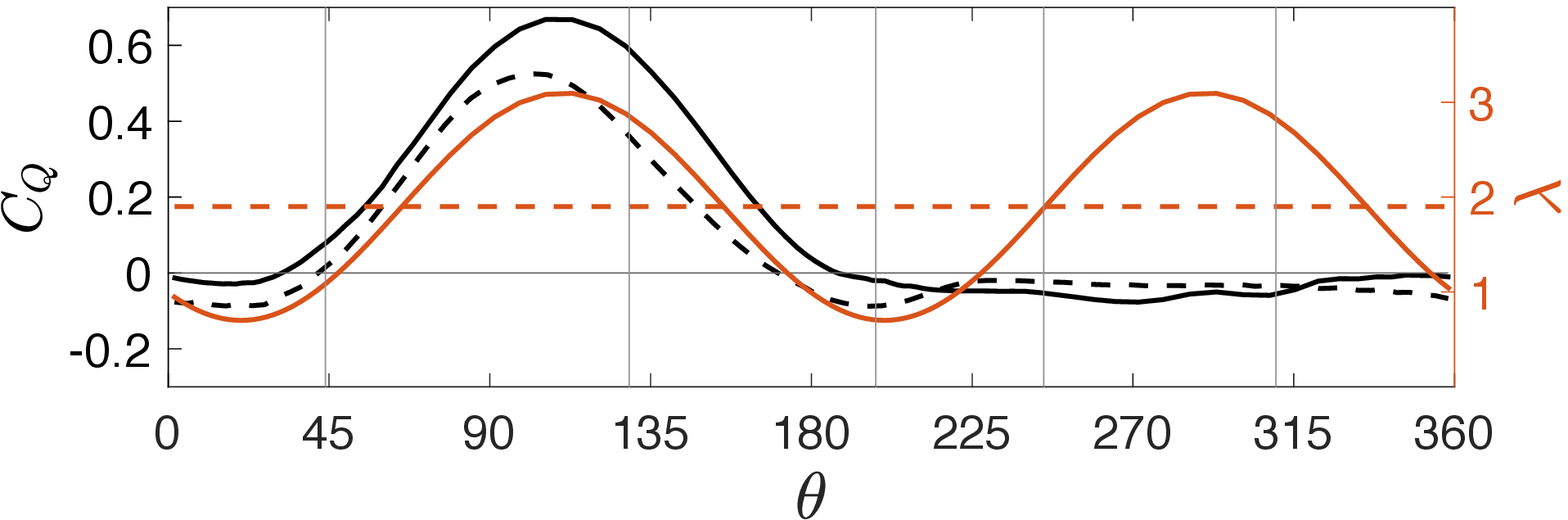}
            \caption{Torque coefficient (black; left axis) and \protect\linebreak tip speed ratio (red; right axis)} 
            \label{fig:intra_b}
        \end{subfigure}
    \end{minipage}
    \hfill
    \begin{minipage}{0.49\textwidth}%
        \begin{subfigure}{\linewidth}%
            \includegraphics[width=\linewidth]{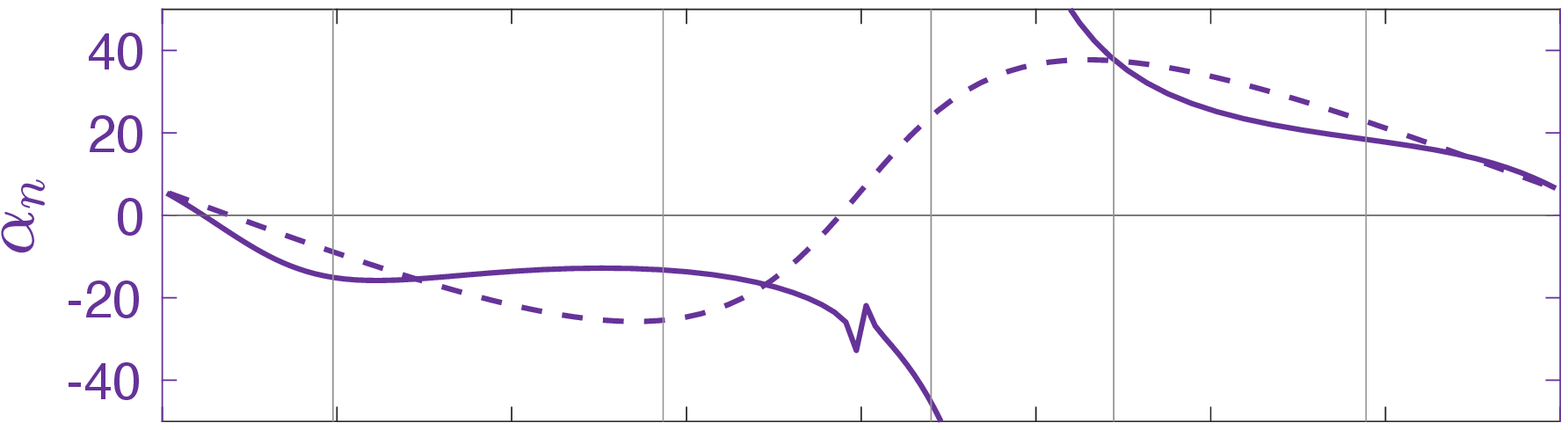}
            \caption{Nominal angle of attack} 
            \label{fig:intra_c}
        \end{subfigure}
        \begin{subfigure}{\linewidth}%
            \includegraphics[width=\linewidth]{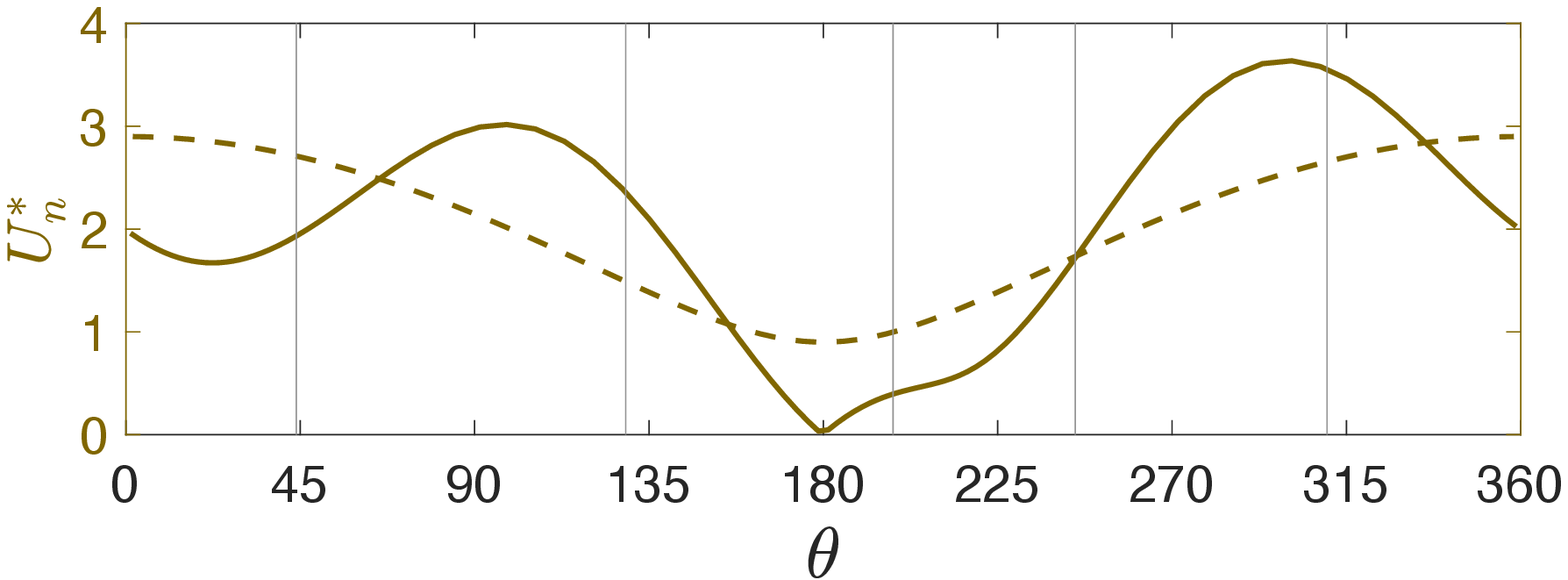}
            \caption{Relative flow velocity} 
            \label{fig:intra_d}
        \end{subfigure}
    \end{minipage}
    \caption{Analysis through a single blade rotation for sinusoidal TSR (solid lines) compared with constant optimal TSR (dashed lines).}
    \label{fig:pmd_all}
\end{figure}

\begin{figure}
    \centering
    \includegraphics[width=0.72\linewidth]{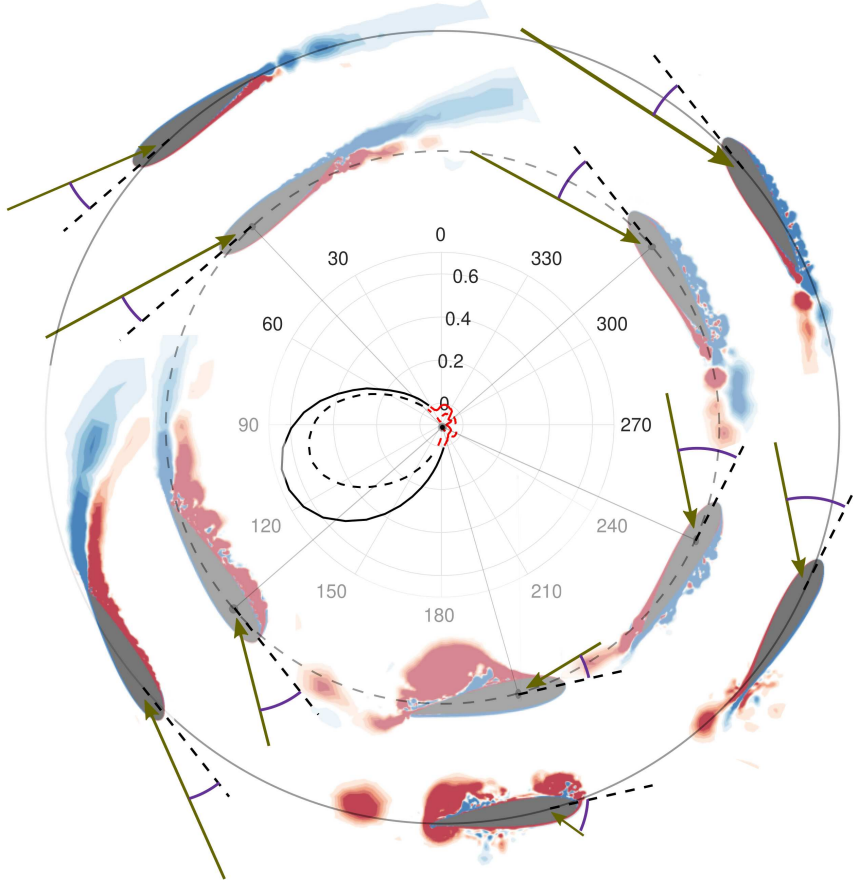}
    \caption{Span-averaged vorticity fields compared between sinusoidal TSR (outer images) and the constant optimal TSR (inner images) at azimuthal positions marked in Fig.~\ref{fig:pmd_all}. Green vectors and purple angles represent the nominal relative flow.
    Torque coefficient $C_Q$ is reproduced in polar form with the negative values in red color.}
    \label{fig:pmd_fields}
\end{figure}

\red{For this turbine geometry, rotation at $\lambda=1.9$ provides the peak performance under constant angular velocity conditions. However, by prescribing intracycle control, or a sinusoidal angular velocity profile (with $\overline{\lambda}=1.9$), the turbine power is enhanced by 40\% \cite{Dave2021intracycle,Strom2017} as displayed in Fig.~\ref{fig:pmd_all}a. The mechanism for power enhancement can be explained by Fig.~\ref{fig:pmd_all}b, where the maximum angular velocity (peak $\lambda$) aligns with the peak in torque, enhancing the power generation. The dynamic stall process is modified due to the changes in angle of attack and relative flow velocity shown in Fig. \ref{fig:pmd_all}c and Fig. \ref{fig:pmd_all}d.} 

One of the prominent features of intracycle control is that the nominal angle of attack rapidly rises to $18\de$ and remains roughly constant for most of the power generating stroke, from $\theta=45\de$ to $135\de$. Figure \ref{fig:pmd_fields} contrasts the nominal relative flow variation for both simulations, superimposed with span-averaged instantaneous vorticity fields. The higher relative flow velocity and lower angle of attack reduces flow separation and delays LEV formation for sinusoidal TSR. \red{This is consistent with the delay and increase in the peak torque generation in Fig.~\ref{fig:pmd_all}b.} 

At $\theta=200\de$, the sinusoidal TSR has ended its power production cycle and produces a trailing edge vortex that sheds the bound vorticity as the angle of attack switches sign, \red{causing an inverse relative flow on the outside of the blade.} 
\red{The remaining portion of the cycle is marked by recovery and reattachment. For the sinusoidal TSR,} minimal vorticity has developed at $\theta=245\de$ due to a history of low relative flow and drastic angle of attack transition, despite a nominal $U_n^*$ identical to that of optimal TSR. Proceeding downstream beyond $\theta=300\de$, the outside surface experiences partial separation similar to that for optimal TSR.

\begin{figure}
    \centering
    \includegraphics[width=\linewidth]{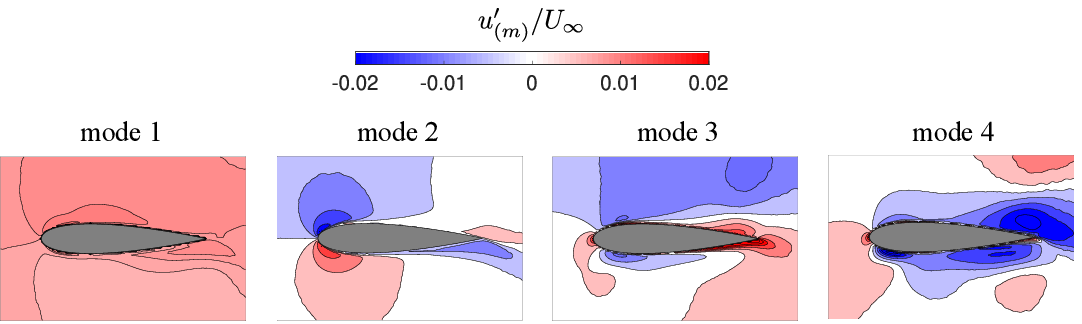}
    \caption{Visualization of first four modes with $x$-velocity contours for sinusoidal TSR.}
    \label{fig:pmd_modes_u}
\end{figure}

\begin{figure}
    \centering
    \includegraphics[width=0.49\linewidth]{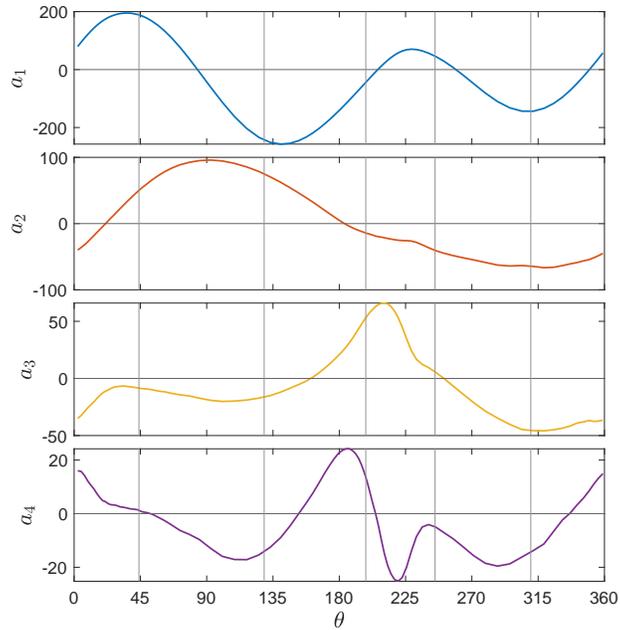}
    \caption{Time development coefficients of modes 1-4 for sinusoidal TSR.}
    \label{fig:intra_e}
\end{figure}

\red{Fig.~\ref{fig:pmd_modes_u} displays the first four modes from the sinusoidal TSR flow, which capture 98.6\% of the energy. The first two modes are qualitatively similar to those of the optimal TSR in Fig.~\ref{fig:const_modes_u}. However modes 3 and 4 display more complexity by exhibiting flow separation features on both the inside (bottom) and outside (top) of the blade.}

\red{The corresponding mode coefficients for sinusoidal TSR are provided in Fig.~\ref{fig:intra_e}.  Qualitatively these can be compared to the mode coefficients for optimal TSR in Fig.~\ref{fig:const_all_1_9_d}. The kinematic discrepancies between the optimal TSR and sinusoidal TSR are highlighted by the change in $\lambda$ (Fig.~\ref{fig:pmd_all}b), resulting in a significant rate of change in the nominal relative flow (Fig.~\ref{fig:pmd_all}c and \ref{fig:pmd_all}d). }While $U_n^*$ is in phase with the angular velocity variation, the magnitude reaches zero at $\theta=180\de$. Notably, mode 1 imitates $U_n^*$ along with reduced magnitudes on the downstream side but a phase shift of $45\de$ to $70\de$ is observed between its coefficient and $U_n^*$. \red{The activation of mode 2 has similar trends to the optimal TSR flow, peaking with max power generation.} 

\red{Modes 3 and 4, representing detached flow, are activated later in the cycle for the sinusoidal TSR, at or near the minimum point for relative velocity at $\theta=180$ deg.} 
The distinct flow separation manifests prominently in the coefficient for mode 4. As mode 4 contains significant reverse flow on the inner surface, its positive values from $\theta=140\de$ to $205\de$ represent the flow separation and LEV formation for optimal TSR. \red{In contrast, mode 4 activation is delayed for sinusoidal TSR, positive between $\theta=155\de$ to $210\de$, with less relative magnitude as compared to mode 1. } 

\section{\label{sec:discussion}Discussion}

\subsection{Flow curvature effects}

\begin{figure}
    \centering
    \includegraphics[width=0.4\linewidth]{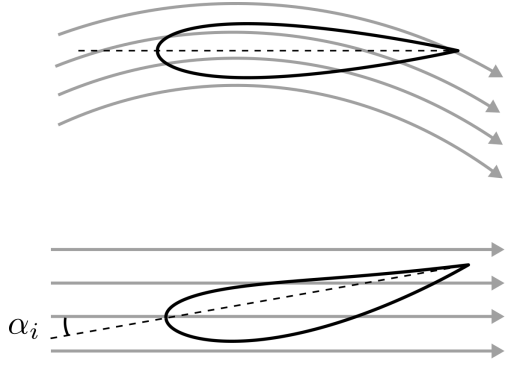}  \caption{Demonstration of virtual camber and incidence due to flow curvature.}
    \label{fig:camber_schem}
\end{figure}

\begin{figure}
    \centering
    \begin{subfigure}[t]{\columnwidth}
        \centering
        \includegraphics[width=0.42\linewidth]{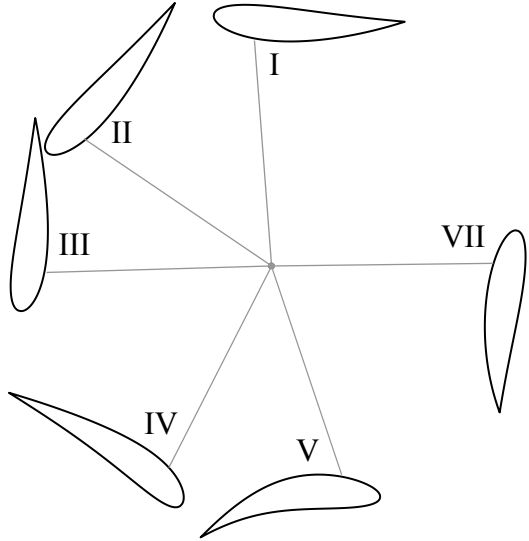}  \caption{Virtual camber variation corresponding to the marked positions in Fig.~\ref{fig:const_all_1_9}. Chord-to-radius ratio is not to scale.} \vspace{24pt}
        \label{subfig:camber_1_9_all}
    \end{subfigure} 
    \begin{subfigure}[t]{0.45\columnwidth}
        \centering
        \includegraphics[width=\linewidth]{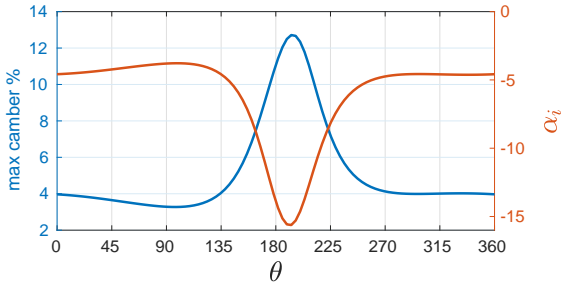}
        \caption{Maximum camber and incidence angle.}
        \label{subfig:camber_1_9a}
    \end{subfigure} \hspace{2mm}
    \begin{subfigure}[t]{0.42\columnwidth}
        \centering
        \includegraphics[width=\linewidth]{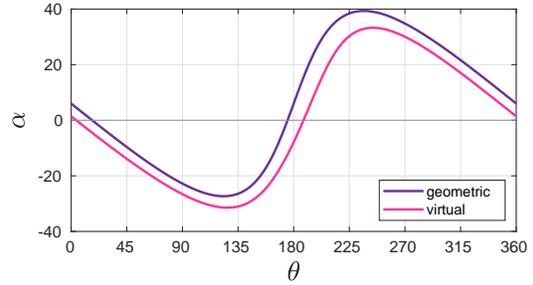}
        \caption{Shift in angle of attack.}
        \label{subfig:camber_1_9b}
    \end{subfigure} 
    \caption{Flow curvature effects - virtual transformation of the foil computed as per \cite{Migliore1980} for $\lambda=1.9$, $c/R=0.47$, and mounted at the quarter chord location.}
    \label{fig:camber_1_9}
\end{figure}

One of the dominant effects of rotation is the relative flow variation along the chord length of a CFT foil, especially for the relatively high chord-to-radius ratio of 0.47 simulated.
The streamlines of relative flow in the blade reference frame are concentric circles, where the center revolves around the turbine center as a function of the blade's azimuthal position. Migliore et al.~\cite{Migliore1980} perform a conformal mapping of this relative flow to determine an equivalent virtually cambered blade in rectilinear flow as demonstrated in Fig.~\ref{fig:camber_schem}. This transformation also results in a shift in the incidence angle, $\alpha_i$, from the geometric angle of attack, $\alpha_n$. This mapping is visually depicted in Fig.~\ref{fig:camber_1_9} for the flow at optimal TSR. The camber and incidence angle follow the trend of the rate of change in $\alpha_n$ though it is not directly proportional.
The theoretical foil transformation does not account for the gradient of relative velocity along the turbine radius or the \red{induced flow} that includes the altered oncoming flow on the downstream side (due to the wake of the opposing foil). 
Nevertheless, it allows visualizing and predicting the qualitative effect of flow curvature on the outcome, particularly on the upstream half of the cycle.

The effect of flow curvature manifests in the non-zero fluctuating pitching moment. During the upstream sweep for optimal TSR, the virtual camber is positive (relative to the angle of attack) which is expected to shift the center of pressure towards the trailing edge and hence partly explains the increasing CW moment between positions I and II.
The camber increases rapidly starting at $\theta \approx 130\de$, and by phase V, the theoretical camber reaches its maximum (Fig.~\ref{subfig:camber_1_9a}). However, its effect is difficult to assess since the blade travels through low relative flow at this point in the cycle.
As the blade proceeds beyond $\theta=225\de$, the angles of attack are lower than the nominal values due to slower oncoming flow. The incidence angle shift in Fig.~\ref{subfig:camber_1_9b} also contributes to the discrepancy between geometric and actual angles of attack, which prevents fully separated flow. The relative flow incidence is  adequately high to cause separation and hence loss of suction pressure on the trailing portion of the outer surface. 
It is hypothesized that the negative virtual camber relative to angle of attack exacerbates this separation behavior.

\begin{figure}
    \centering
    \begin{subfigure}{0.45\columnwidth}
        \centering
        \includegraphics[width=\linewidth]{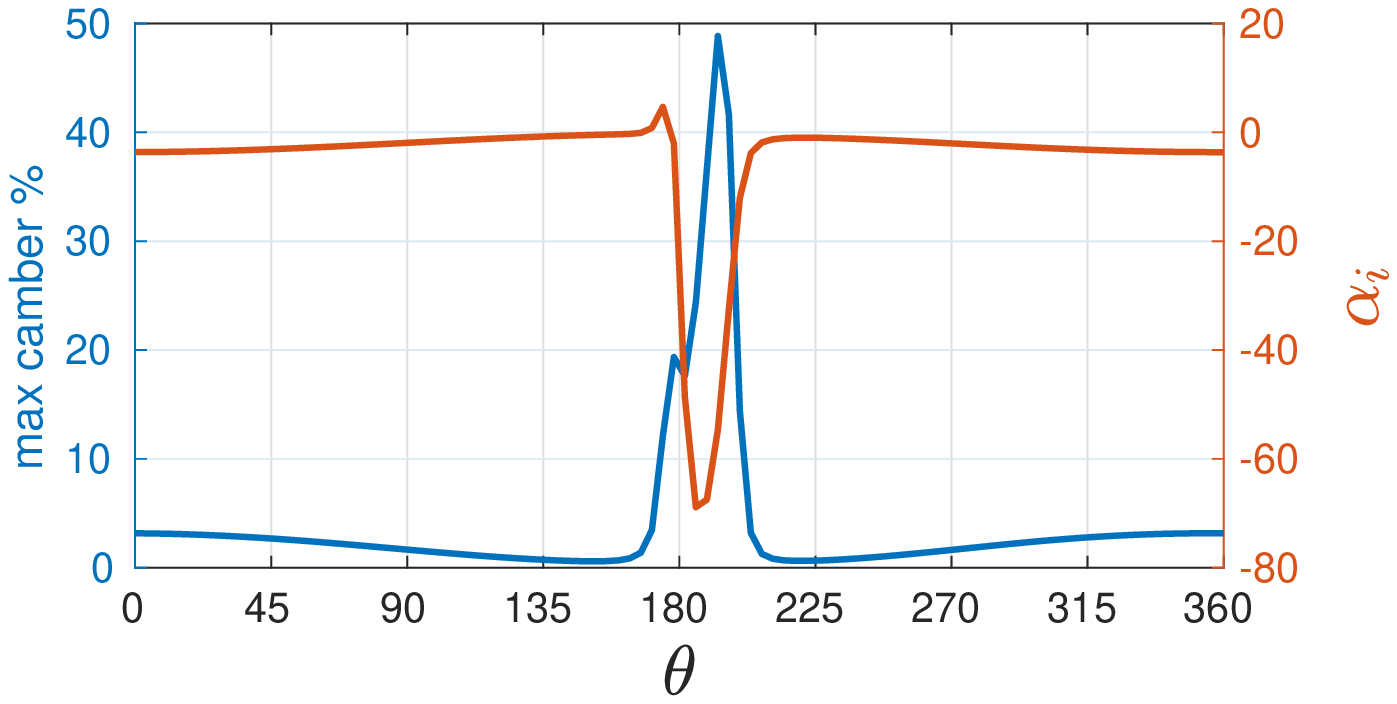}
        \caption{sub-optimal TSR}
        \label{subfig:camber_1_1}
    \end{subfigure} \hspace{2mm}
    \begin{subfigure}{0.45\columnwidth}
        \centering
        \includegraphics[width=\linewidth]{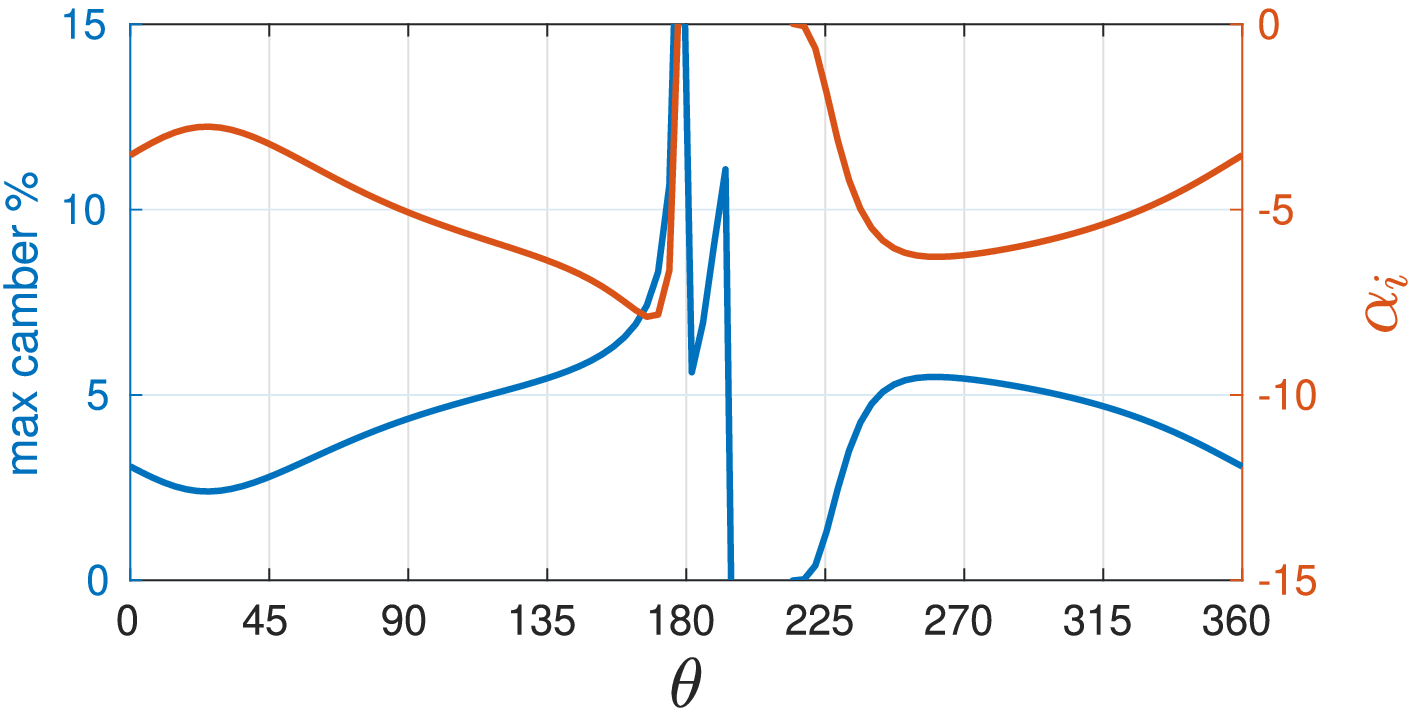}
        \caption{sinusoidal TSR}
        \label{subfig:camber_intra}
    \end{subfigure} 
    \caption{Flow curvature effects - theoretically calculated maximum camber and incidence angle for the sub-optimal TSR and sinusoidal TSR.}
    \label{fig:camber_other}
\end{figure}

The calculated virtual camber and incidence for sub-optimal TSR and the sinusoidal TSR are shown in Fig.~\ref{fig:camber_other}. Both these kinematics display higher camber values around $\theta=180\de$ due to a high rate of change of $\alpha_n$.
For the low tip speed ratio, large \red{induced flow} effects dominate at this position and hence the virtual camber and incidence are expected to have a minimal impact.
While the blade aerodynamic loads are not analyzed for the sinusoidal TSR, the virtual camber is expected to enhance the CW moment during the longer power generating attached-flow phase as observed for optimal TSR.
However at the end of the upstream motion, as the relative flow tends towards zero and the angle of attack reverses signs, the effect of the large theoretical camber would be difficult to isolate.

\subsection{History effects and \red{induced flow}}

History or dynamic effects refers to the flow state at a specific instance being a function of the system trajectory rather than simply the conditions imposed at that instance. In contrast, \red{induced flow refers to the prior field produced by the blades themselves affecting the instantaneous state.} The interaction between these two phenomena, and their effect on separation dynamics, are highlighted in this section.

For optimal TSR, the positive pressure during reattachment of the boundary layer at phases II and III 
contributes to the CW moment. This phenomenon is similar to that observed in pitching foils, causing the normal force and pitching moment to surpass their static values for the minimum angle of attack during the pitch-down and reattachment \cite{Carr1988,Mulleners2012}. The downstream stroke also experiences reattachment on the inside portion of the blade (phase VI in Fig.~\ref{fig:const_all_1_9}), but has negligible pressure as a consequence of the low relative velocity field.

Phases V to I in Fig.~\ref{fig:const_all_1_9} reveal that LEV formation or fully separated flow is prevented on the outside due to angles of attack lower than the nominal/geometric values. Notably, a fully attached flow with considerable lift generation similar to phase I-II on the upstream side is also never seen on the downstream side. This is a result of \red{induced flow} since the low angle of attack regime in phase V, which would be expected to produce attached flow, has the foil traveling through its own wake through minimal relative flow velocity. 

\red{Induced flow} plays a pivotal role for the sub-optimal TSR rotation as outlined in section \ref{sec:results}B, albeit in a distinct manner from the optimal TSR rotation. The stall vortex is larger and stronger and detaches early on in the cycle, impacting the blade loads through a longer span of time. The leading and trailing edge vortices cause transient fluctuations in the aerodynamic loads and create additional vortices on the outer surface of the blade, immediately affecting the same blade that they emanate from. In addition, different separation behavior in terms of the vortex sizes and strengths between the upstream and downstream sweeps is seen in Fig.~\ref{fig:const_all_1_1} due to the \red{induced} low-velocity effects from the opposite blade as observed for the higher $\lambda$.

\subsection{Comparison with non-rotating foils}

\begin{figure}
    \centering
    \includegraphics[width=0.5\linewidth]{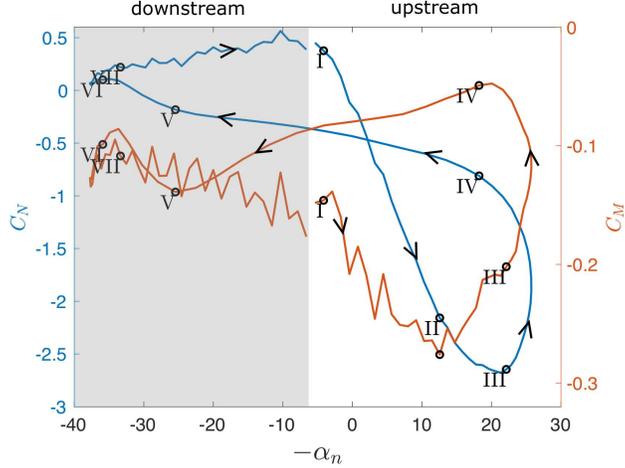}
    \caption{Normal force and pitching moment vs nominal angle of attack for optimal TSR.}
    \label{fig:const_aoa}
\end{figure}

The aerodynamic loads are plotted in terms of the instantaneous nominal angle of attack in Fig.~\ref{fig:const_aoa} to produce a conventional stall loop for the optimal TSR. The grey area represents downstream rotation or the recovery stroke of the turbine in which the nominal angle of attack significantly misrepresents the true relative flow. While the nominal value also has inaccuracies on the upstream stroke, it is expected to more closely follow the true relative flow and hence is utilized in this section to compare the dynamic stall loop with prior research.

The stall cycle for the normal force creates the typical figure-eight, though the values are negative-biased due to the preset pitch angle of 6 degrees. Furthermore, the downstream greyed area is not a mirror image of the upstream cycle despite the nominal relative flow indicating such a profile (due to history and \red{induced flow}). 
In contrast to a pitching foil, the drop in normal force at phase III is significantly earlier from the onset of stall at phase IV due to the varying relative velocity among other factors.

As seen in literature on pitching and plunging foils \cite{Mulleners2012,Kim2017,Ribeiro2020}, the detachment of LEV, marked by the peak of the LEV mode, causes a ``pitch-down" moment starting at phase IV.
However, non-rotating blades experience a near-constant pitching moment before the detachment of LEV. 
In contrast, the CFT blade undergoes an increased pitch-down moment before LEV separation due to virtual camber during attached flow and a positive pressure associated with reattachment on the outside. This is followed by a reduced pitch-down moment due to flow separation and LEV formation.
The pitching moment remains CW (pitch-up) on the downstream side because the flow is neither fully attached nor fully separated within this region. 
The effective ``pitch-down motion", that is, the decrease in nominal angle of attack, begins before reaching phase IV as apparent from Fig.~\ref{fig:const_aoa}.
Using the definition of light and deep stall based on stall onset relative to the start of the pitch-down motion \cite{Mulleners2012}, the optimal TSR rotation can be characterized as experiencing light stall. 

For sub-optimal TSR, detachment of the stall vortex occurs much before the start of the ``pitch-down motion" and hence this rotation can be said to be undergoing a deep stall.
However, unlike a pitching or plunging foil, this stall vortex remains in the vicinity of the blade and induces other vortices along with transitional peaks in the pitching moment.
Due to the complex flow physics, the stall events such as the stall onset at phase II or the aerodynamic load peaks at phases III and V in Fig.~\ref{fig:const_all_1_1} do not correspond to a single mode coefficient. 

\section{\label{sec:conclusion}Conclusion}

\red{Three diverse rotational kinematics of a two-bladed cross-flow turbine are simulated using large-eddy simulation for the purpose of investigating the dynamic stall characteristics. The cases include an optimal TSR at $\lambda=1.9$, which is close to the maximum power producing kinematics for the turbine geometry and a sub-optimal TSR at $\lambda=1.1$ in which the turbine blade experiences deep stall.}

\red{Finally, a sinusoidal TSR with $\overline{\lambda}=1.9$ is contrasted directly with the optimal TSR in terms of how the power and vortex dynamics are governed by the relative angle of attack and velocities. By varying the angular velocity intracycle, the velocity peak is aligned with the torque peak, enhancing power by 40\%.  This has an effect of delaying separation and the onset of dynamic stall, but also complicates the dynamic stall vortex formation and convection as the relative angular velocity drops significantly on the stroke reversal. }

\red{Proper orthogonal decomposition of the velocity fields reveal modes whose time development coefficients have trends specific to the critical events in the dynamic stall cycle on the blade.} The two most energetic modes and their time development coefficients primarily represent the relative velocity magnitude and the lift generation. The third and fourth modes account for the stall vortex formation and shedding. \red{These trends are stronger in the optimal and sinusoidal TSR cases as opposed to the sub-optimal TSR where the vortex dynamics are initiated earlier in the cycle and dominate throughout the rotation.}

At the optimal TSR of $\lambda=1.9$, virtual camber due to flow curvature and boundary layer reattachment from the downstream motion of the blade cause a peak in the ``pitch-down" moment before the formation and detachment of the leading edge vortex. These events are captured by the relative significance or peaks in mode coefficients. \red{Induced flow} plays a prominent role during the downstream motion as the blade travels through its own wake and then through low velocity oncoming flow, resulting in an asymmetry from the upstream motion. 

\red{At the sub-optimal TSR of $\lambda=1.1$, the variation in angle of attack and relative velocity is enhanced, causing stronger induced flow.} Blade-vortex interactions are dominant and responsible for fluctuations in the pitching moment.  The earlier and stronger vortex shedding is captured by the velocity-based mode coefficients, and is responsible for lower normal force and moment coefficients. 

\red{The specific forces and power curves reported are dependent on flow conditions and turbine geometry, and may vary across cross-flow turbine configurations. However, the flow analysis illustrates the complexity of modeling a cross-flow turbine and the interdependence of rotational flow with the dynamic stall process. Furthermore, modal decomposition is shown to capture critical points in the dynamic stall cycle, distinguishing between two rotational kinematics, and can serve as a potential guide to designing blade-level active flow control.} 

\begin{acknowledgments}
The authors would like to thank Abigale Snortland, Ari Athair, Owen Williams, Karen Mulleners and S\'{e}bastien Le Fouest for motivating discussions over the topics in this work.
The authors acknowledge financial support from the Advanced Research Projects Agency-Energy (ARPA-E) under award number DE-AR0001441.
\end{acknowledgments}

\bibliography{darrieus,darrieus2,darrieus_extra,mypapers,rom,dynstall}

\end{document}